\documentclass[%
 reprint,
 amsmath,amssymb,
 aps,
 prd,
]{revtex4-1}

\usepackage{xcolor}
\usepackage{graphicx}
\usepackage{dcolumn}
\usepackage{bm}
\usepackage{hyperref}

\def\bea{\begin{eqnarray}}
\def\eea{\end{eqnarray}}

\begin{document}

\title{Newtonian Binding from Lattice Quantum Gravity}
\author{Mingwei Dai}
\affiliation{Department of Physics, Syracuse University, Syracuse, NY 13244}
\author{Jack Laiho}
\affiliation{Department of Physics, Syracuse University, Syracuse, NY 13244}
\author{Marc Schiffer}
\affiliation{Institut f\"{u}r Theoretische Physik, Universit\"{a}t Heidelberg,
Philosophenweg 16, 69120 Heidelberg, Germany}
\author{Judah Unmuth-Yockey}
\email{jfunmuthyockey@gmail.com}
\affiliation{Department of Physics, Syracuse University, Syracuse, NY 13244}
\affiliation{Fermi National Accelerator Laboratory, Batavia, IL 60510}

\date{\today}

\begin{abstract}
We study scalar fields propagating on Euclidean dynamical triangulations (EDT).  In this work we study the interaction of two scalar particles, and we show that in the appropriate limit we recover an interaction compatible with Newton's gravitational potential in four dimensions.  Working in the quenched approximation, we calculate the binding energy of a two-particle bound state, and we study its dependence on the constituent particle mass in the non-relativistic limit.  We find a binding energy compatible with what one expects for the ground state energy by solving the Schr\"{o}dinger equation for Newton's potential.  Agreement with this expectation is obtained in the infinite-volume, continuum limit of the lattice calculation, providing non-trivial evidence that EDT is in fact a theory of gravity in four dimensions.  Furthermore, this result allows us to determine the lattice spacing within an EDT calculation for the first time, and we find that the various lattice spacings are smaller than the Planck length, suggesting that we can achieve a separation of scales and that there is no obstacle to taking a continuum limit.  This lends further support to the asymptotic safety scenario for gravity.

\end{abstract}

\maketitle

\section{Introduction}
\label{sec:intro}

Quantum gravity is one of the great outstanding problems of theoretical physics.  One possible approach to obtaining a consistent, predictive theory is the asymptotic safety scenario of Weinberg \cite{Weinberg:1980gg}.  It is well-known that general relativity is not perturbatively renormalizable \cite{tHooft:1974toh, Goroff:1985th}.  Although a low-energy effective theory can be formulated \cite{Donoghue:1993eb,Donoghue:1994dn}, it comes with an infinite number of unknown couplings and is thus limited in terms of its predictive power.  Asymptotic safety rests on the hope that there is a strongly coupled ultra-violet fixed point, making it effectively renormalizable nonperturbatively, with only a finite number of parameters needing to be specified from experiment.  One would also hope that the number of relevant parameters needing to be specified from experiment is small, since a theory with few input parameters is more predictive than one with many.  If asymptotic safety is realized for gravity, nonperturbative methods will be crucial to convincingly establish this and to explore the detailed properties of the theory. 

Evidence that the asymptotic safety scenario is realized for gravity comes mainly from the lattice \cite{Ambjorn:2005qt,Ambjorn:1998xu,Ambjorn:2008wc,Laiho:2016nlp} and from the functional renormalization group \cite{Wetterich:1992yh,Ellwanger:1993mw,Morris:1993qb}, which was pioneered for gravity in \cite{Reuter:1996cp} (see, e.g., \cite{Niedermaier:2006ns,Percacci:2007sz,Litim:2008tt,Reuter:2012id,Eichhorn:2018yfc,Pereira:2019dbn,Reichert:2020mja,Pawlowski:2020qer,Bonanno:2020bil} for introductions and reviews).
This paper follows up on previous work \cite{Laiho:2016nlp} using Euclidean dynamical triangulations (EDT), which is one of the original approaches to quantum gravity on the lattice \cite{Ambjorn:1991pq, Agishtein:1991cv,Catterall:1994pg}.  EDT is a discretization of Euclidean quantum gravity, where the path-integral is given by a sum over geometries, weighted by the Einstein-Hilbert action plus the action of the matter sector.  In a lattice formulation of an asymptotically safe theory, the ultraviolet fixed point would appear as a continuous phase transition, the approach to which defines a continuum limit.  This is the first test that EDT must pass.  In addition to this, the theory must reproduce classical general relativity in the appropriate limit.  If EDT could pass these two tests, it would be a candidate for an ultraviolet-complete theory of four dimensional quantum gravity. 

 We briefly review the evidence so far that EDT may realize the asymptotic safety scenario for gravity.
 Ref.~\cite{Laiho:2016nlp} introduced a fine tuning of the exponent of a local measure term, showing that the tuning of this parameter is necessary in order to recover semiclassical physics.  Although the local measure term was first introduced some time ago in Ref.~\cite{Bruegmann:1992jk}, the nice features of this tuning have only been appreciated recently.  Once the tuning is done, four-dimensional geometries are recovered that resemble Euclidean de Sitter space.  There also appears to be no obstacle to taking the continuum limit by following the first order line to a possible critical endpoint, and ensembles following this procedure were generated at a number of different lattice spacings.  The global Hausdorff dimension was measured using finite-volume scaling and shown to be close to four \cite{Laiho:2016nlp}.  The spectral dimension, which is a fractal dimension defined by a diffusion process, varies with distance scale and approaches a value close to four at long distances.  The variation of the spectral dimension with distance had also been found earlier in other approaches \cite{Ambjorn:2005db, Lauscher:2005qz, Carlip:2017eud}.  The average over geometries in Ref.~\cite{Laiho:2016nlp} gives a result that is close to that of Euclidean de Sitter space, and the quantitative agreement with the classical solution gets better as the proposed continuum limit is approached.  The agreement between the classical solution and the lattice data is actually the worst at long distances, but improves as the lattice spacing is reduced.  This might seem surprising, since it is typically the short-distance behavior that is modified by discretization effects, but this type of effect on long-distance behavior is typical when a symmetry of a theory is broken by the regulator, for example by the finite lattice spacing in the case of lattice regularization.  Then a fine-tuning is needed to approximately restore the symmetry at finite lattice spacing.  Residual symmetry-breaking effects appear at finite lattice spacing and can modify the physics that depends on the symmetry.  The correct long-distance physics is then only recovered when the symmetry is restored in the continuum limit.  An example of this is the Wilson fermion formulation of lattice quantum chromodynamics, where the lattice regulator breaks chiral symmetry.  There a fine-tuning is required to restore chiral symmetry, and even then, at finite lattice spacing the chiral symmetry breaking leads to distortions of the pion sector, which contains the lightest physical states of the theory.  Ref.~\cite{Laiho:2016nlp} argued by analogy to the Wilson fermion case that the symmetry that is broken by dynamical triangulations is continuum diffeomorphism invariance. 
 
 Further evidence that dynamical triangulations recovers the correct long-distance limit was given in Ref.~\cite{Catterall:2018dns}, where it was shown how to incorporate K\"{a}hler-Dirac fermions \cite{Kahler:1962}.  This approach generalizes the staggered fermion formulation to the random lattices of dynamical triangulations without the need to introduce vielbeins or spin connections.  It is well known that the  K\"{a}hler-Dirac action reduces to four copies of Dirac fermions in the flat-space, continuum limit \cite{Banks1982}.  Ref.~\cite{Catterall:2018dns} found evidence that this is the case for K\"{a}hler-Dirac fermions on dynamical triangulations in the large-volume (small curvature) limit, but only if the continuum limit is also taken.  This is seen in the approximate four-fold degeneracy in the low-lying eigenvalues of the K\"{a}hler-Dirac matrix, and in the degeneracy of scalar bound states in the continuum limit.  The four-fold degeneracy is lifted by lattice effects in a similar manner to what is found with staggered fermions in lattice quantum chromodynamics (QCD) \cite{Bazavov:2009bb}, but as in lattice QCD, the degeneracy appears to be restored in the continuum limit.  The evidence that these discretization effects vanish suggests that continuum K\"{a}hler-Dirac fermions in four dimensions are recovered at that point.  An additional advantage of K\"{a}hler-Dirac fermions is that they possess an exact $U(1)$ symmetry, which is related to continuum chiral symmetry.  A study of fermion bilinear condensates provides strong evidence that this $U(1)$ symmetry is not spontaneously broken at order the Planck scale, implying that fermion bound states do not acquire unacceptably large masses due to chiral symmetry breaking.  These results for K\"{a}hler-Dirac fermions in EDT are highly non-trivial and provide further evidence for the asymptotic safety scenario for gravity and matter.
 
 In this work we continue our investigation of matter fields living on dynamical triangulations, this time focusing on scalar fields.  We follow the work of de Bakker and Smit \cite{deBakker:1996qf} in our implementation of scalar fields.  They looked at scalars propagating on EDT backgrounds (without the inclusion of a local measure term) and the binding energy of bound states of scalar particles.  They found clear evidence of gravitational binding, such that there was an attractive force between the scalars, but they were not able to make contact with the Newtonian limit (see \cite{Smit:2021lyq} for recent work attempting to understand the systematic errors associated with the original de Bakker, Smit analysis). In the original reference \cite{deBakker:1996qf}, these authors outlined a prescription for recovering the Newtonian limit.  In this work we follow their prescription, using our approach to taking the continuum limit and our general strategy, which has been successful for other observables.  We confirm the finding that there is an attractive force between scalar particles, and after taking the continuum, infinite volume limit, we find results compatible with Newtonian gravity in the non-relativistic limit.  We can therefore use Newton's law to infer a value of Newton's constant $G$, which allows us to convert lattice units into units of the Planck length.  We find that our lattice spacings are smaller than the Planck length and that for our finest lattice spacings we are starting to see a separation of scales, such that the lattice spacing is starting to become much smaller than the Planck scale.  This provides evidence that not only does the formulation recover the correct long-distance physics, but also that there is no barrier to taking the continuum limit.
 
 This paper is organized as follows:  Section \ref{sec:edt} reviews the EDT formulation and the details of the lattice simulations.  Section \ref{sec:theo-back} reviews how to add scalar fields to EDT and how to compute the scalar propagator and the binding energy of a bound state of two scalar particles.  Section \ref{num-res} presents our numerical results for the binding energy and the analysis needed to recover the form of the interaction and Newton's constant.  We conclude in Section \ref{sec:conclusion}.  Finally, an appendix gives some details of the fitter used in the analysis.

\section{Euclidean dynamical triangulations}
\label{sec:edt}

\subsection{The model}

In the continuum, for fixed global topology, the four-dimensional Euclidean quantum gravity partition function is given by the path integral sum over all geometries, weighted by the Einstein-Hilbert action and the action for the matter sector,
\bea\label{eq:part} Z_E= \int {\cal D}[g] {\cal D}[\phi] e^{-S_{EH}[g]-S_{M}[\phi]},
\eea
where the Euclidean Einstein-Hilbert action is
\bea\label{eq:ERcont} S_{EH}=-\frac{1}{16\pi G}\int d^4x\sqrt{g}(R -2\Lambda),
\eea
with $R$ the curvature scalar, $\Lambda$ the cosmological constant, $G$ Newton's constant, and we adopt for the matter sector a real, non-interacting massive scalar field minimally coupled to gravity,
\bea 
\label{eq:matter-action}
S_{M}=\int d^4x\sqrt{g}\left(\frac{1}{2}g^{\mu\nu}\partial_\mu \phi \partial_\nu \phi
+ \frac{1}{2}m_{0}^2 \phi^2 \right),
\eea
with bare mass $m_0$ for the scalar field.

The partition function for lattice quantum gravity that we use in this work is that of Euclidean dynamical triangulations, where the path integral at finite lattice spacing is approximated by the sum over all four-geometries constructed by gluing together four-dimensional, equilateral simplices. It is given by \cite{Ambjorn:1991pq, Bilke:1998vj}
\bea\label{eq:Z} Z_E = \sum_T \frac{1}{C_T}\left[\prod_{j=1}^{N_2}{\cal O}(t_j)^\beta\right]e^{-S_{ER}}
\eea
where the factor $C_T$ divides out equivalent ways of labeling the vertices in a given geometry, the term in brackets is a local measure term with the product over all triangles, and ${\cal O}(t_j)$ is the order of triangle $j$, i.e. the number of four-simplices to which it belongs.  $S_{ER}$ is the Einstein-Regge action \cite{Regge:1961px} of discretized gravity,
 \begin{equation} \label{eq:GeneralEinstein-ReggeAction}
S_{ER}=-\kappa\sum_{j=1}^{N_2} V_{2}\delta_j+\lambda\sum_{j=1}^{N_4} V_{4},
\end{equation}
where $\kappa=(8\pi G)^{-1}$, $\lambda=\kappa\Lambda$, $\delta_j=2\pi-{\cal O}(t_j)\rm{arccos}(1/4)$ is the deficit angle around a triangular hinge $t_j$, and where the volume of a $d$-simplex of equilateral edge length $a$ is given by
\begin{equation} \label{eq:SimplexVolume}
V_{d}=\frac{\sqrt{d+1}}{d!\sqrt{2^{d}}}a^d.
\end{equation}

It is standard to absorb the overall numerical factors into constants and to perform the sums in Eq.~(\ref{eq:GeneralEinstein-ReggeAction}) so that the lattice action is given by the concise form
\bea\label{eq:ER}  S_{ER}=-\kappa_2 N_2+\kappa_4N_4,
\eea
with $N_4$ the number of four simplices and $N_2$ the number of triangles.  The parameters $\kappa_2$, $\kappa_4$, and $\beta$ are inputs to the simulations whose values must be adjusted in order to approach the continuum limit.  We do not include the matter action in the Boltzmann weight when generating our ensembles; this is known as the quenched approximation.  Although this is an uncontrolled approximation, we still expect many features of the full theory to be preserved in the quenched theory.  The main advantage of quenching is that it allows us to reuse the ensembles generated and analyzed in previous works \cite{Laiho:2016nlp}.

\subsection{Simulation details}

The details of the generation of the lattice configurations used in this work are given in Ref.~\cite{Laiho:2016nlp}, and they are summarized briefly here.  The lattice geometries are made by gluing together four-dimensional simplices along their three-dimensional faces.  The four-simplices are equilateral, with constant edge length $a$, and the dynamics is encoded in the connectivity of the simplices.  We sum over a class of triangulations known as degenerate triangulations \cite{Bilke:1998bn}.  This class of triangulations does not obey the combinatorial manifold constraints, though it does lead to a reduction of finite-size effects by approximately a factor of 10 over combinatorial triangulations, giving it some numerical advantages \cite{Bilke:1998bn}.  When the triangulations are degenerate, not all of the neighbors of a given four-simplex are necessarily distinct. Also, distinct four-simplices may share the same five vertex labels.  Such configurations are not allowed when the lattices obey the combinatorial manifold constraints.  Evidence for the existence of a continuum limit for degenerate triangulations with a non-trivial measure term was given in Ref.~\cite{Laiho:2016nlp}.  It is likely that this universality class would be shared by combinatorial triangulations if the continuum limit does in fact exist, as discussed in Ref.~\cite{Laiho:2016nlp}.  

The algorithm for evaluating the partition function in Eq.~(\ref{eq:Z}) is now standard, and consists of a set of ergodic local moves, known as the Pachner moves, which are used to update the geometries \cite{Agishtein:1991cv, Gross:1991je, Catterall:1994sf}.  The proposed local changes are then accepted or rejected using a Metropolis step.  Most of the lattices used in this work were generated using a parallel variant of this standard algorithm known as parallel rejection \cite{Laiho:2016nlp}.  Parallel rejection partially compensates for the low acceptance of the proposed local changes by the Metropolis accept/reject step.  We define a sweep as a fixed number of attempted moves, though the precise number varies across ensembles. A sweep could be $10^8$ or $10^9$ attempted moves, with the acceptance rate depending strongly on the location in the phase diagram. Acceptance rates range from around $10^{-2}$ to $10^{-4}$.  Measurements are made after 10 to 50 sweeps, and our longer runs were for tens of thousands, or even one hundred thousand sweeps.

The global topology of the geometries is fixed to $S^4$. The topology remains fixed since the local Pachner moves are topology preserving, and we choose $S^4$ by starting from the minimal four-sphere at the beginning of the Monte Carlo evolution.  In order to take the infinite lattice volume limit it is necessary to fine-tune the bare parameter $\kappa_4$ to its critical value.  This amounts to a tuning of the bare cosmological constant.  Although this tuning is required to take the infinite lattice volume limit, it does not guarantee that we can take the infinite physical volume limit. The infinite lattice volume limit can be taken even in an unphysical phase where the extent of the lattice ``universe" is only a few lattice spacings, so that the extent of the physical volume is on the order of the ultraviolet cut-off scale.  As shown in Ref.~\cite{Laiho:2016nlp}, there is a region of the phase diagram with extended geometries, so that the large lattice-volume limit also corresponds to the limit of large physical volume.

The simulations are done at approximately fixed lattice four-volumes.  The Pachner moves require the volume to vary in order for them to be ergotic, though in practice the volume fluctuations about the target four-volume are constrained to be small.  This is done by introducing a volume preserving term in the action $\delta \lambda|N_4^f-N_4|$, which keeps the four-volume close to a fiducial value $N_4^f$.  This term does not alter the action when $N_4=N_4^f$, but it does keep the volume fluctuations from becoming too large to be practical in numerical simulations.  In principal one should take the limit in which $\delta \lambda$ goes to zero, but in practice setting it small enough does not appear to lead to significant systematic errors.  In Ref.~\cite{Laiho:2016nlp} the parameter $\delta \lambda$ was taken to be 0.04, though taking this value smaller did not lead to a noticeable change in the results.

\begin{figure}
\begin{center}
\includegraphics[scale=.55]{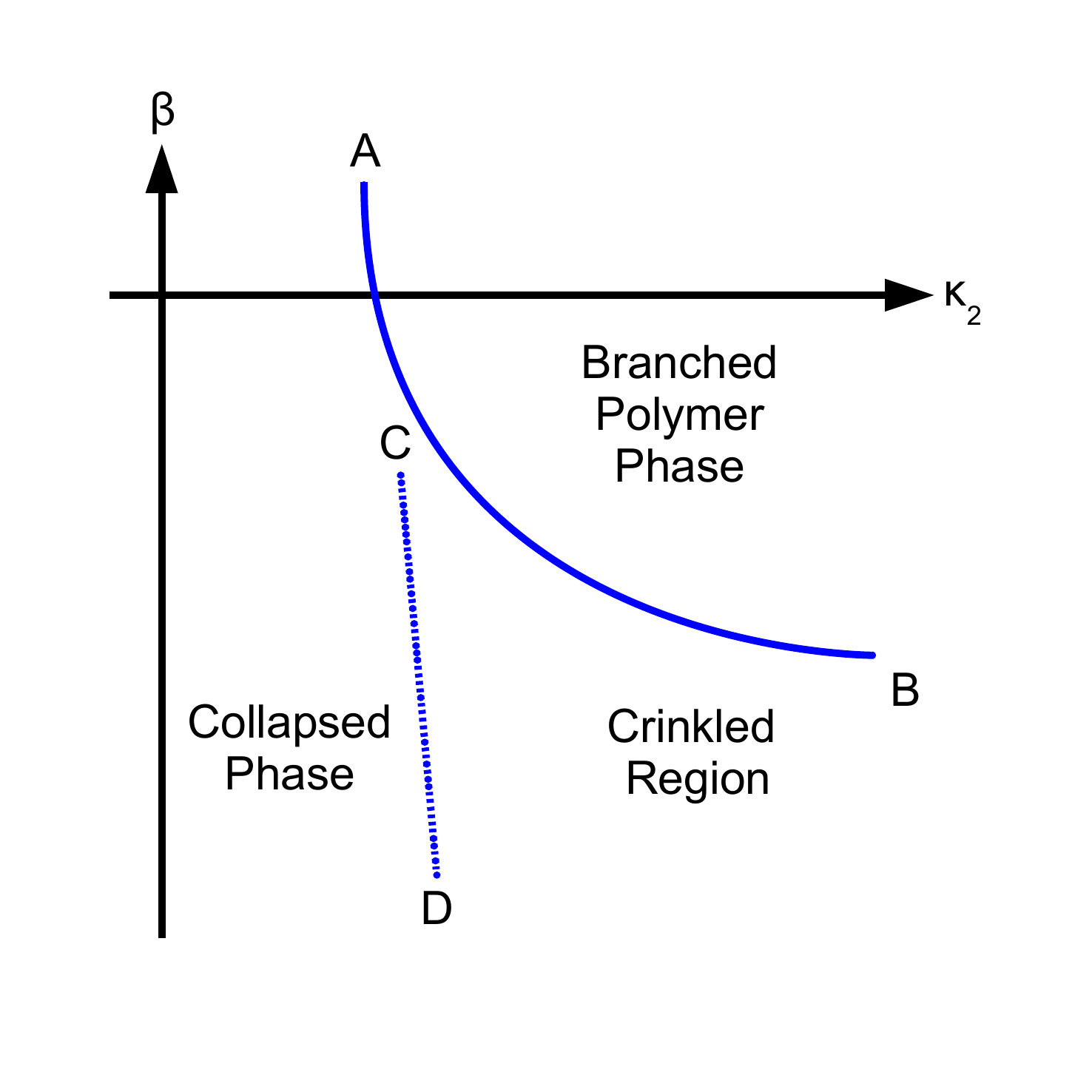}
\vspace{-3mm}
\caption{Schematic of the phase diagram as a function of $\kappa_2$ and $\beta$. \label{fig:phase1}}
\end{center}
\end{figure}

  The phase diagram for the theory is shown in Figure~\ref{fig:phase1}.  The solid line $AB$ is a first order phase transition line that separates the branched polymer phase and the collapsed phase.  Neither of these phases looks much like semiclassical gravity.  The branched polymer phase has Hausdorff dimension 2, while the collapsed phase has a large, possibly infinite, dimension.  The crinkled region does not appear to be distinct from the collapsed phase, but is connected to it by an analytic cross-over.  The crinkled region requires larger volumes to see the characteristic behavior of the collapsed phase, suggesting that it is a part of the collapsed phase with especially large finite-size effects \cite{Coumbe:2014nea}. The results of Ref.~\cite{Ambjorn:2013eha} for combinatorial triangulations are in broad agreement with this picture of the phase diagram.  

The exponent $\beta$ associated with the local measure term must be fine-tuned in order to obtain ensembles with four-dimensional, semiclassical properties.  It was shown in Ref.~\cite{Laiho:2016nlp} that close to the transition line $AB$ the lattice geometries have semiclassical properties, with global Hausdorff dimension close to four and a spectral dimension that approaches four at long distances.  The prescription for recovering semiclassical geometries is to approach the line $AB$ from the left.  Since a first order transition is characterized by tunneling between meta-stable states, if we simulate too close to the transition line then some of the time a given run will be in the wrong state.  In order to minimize contamination from the wrong phase, we only take those parts of a run that are in the correct phase.  We find that this selection procedure becomes unnecessary at the finest lattice spacing considered in this work, since we can simulate somewhat farther from the transition line without losing the four-dimensional finite-volume scaling.  This is fortunate, since future runs will be at ever finer lattice spacings, so will likely not suffer from this systematic error.

Table~\ref{tab:ensembles} shows the parameters for the ensembles used in this work.  Most of these ensembles were generated and discussed in Ref.~\cite{Laiho:2016nlp}, but some of them are new.  We have added a larger lattice volume at our finest lattice spacing, as well as larger and smaller lattice volumes at $\beta=0$ to aid with the infinite-volume extrapolation.  

The determination of the relative lattice spacing follows the procedure of Ref.~\cite{Laiho:2016nlp}, which looked at the return probability of a diffusion process on the lattice geometries.  The return probability is dimensionless, but varies as a function of the diffusion time step, which is not.  One can rescale the diffusion time step so that the return probability lies on a universal curve; the rescaling factor then tells us the relative lattice spacing.  The relative lattice spacings quoted here differ slightly from those first reported in Ref.~\cite{Laiho:2016nlp}.  In that work the overlap of the return probability curves was obtained without regard for the lattice volumes used in the comparison.  Here we match in a self-consistent way ensembles with lattices as close to the same physical volumes as possible.  The shifts are within previously quoted errors, but the central values quoted here are likely more accurate.  The errors quoted here are similar in size to those previously obtained and reflect the uncertainties in matching the curves in this procedure.

\begin{table}
\begin{center}
\begin{tabular}{ccccc}
\hline \hline
\ \ $\ell_{\rm rel}$ \ \ & \ \ $\beta$ \ \ \ & \ $\kappa_2$ \ & \ \ \ \ \ $N_4$ \ \ & \ \ \ Number of configurations \\
\hline
1.59(10) & 1.5 & 0.5886 & \ \ 4000 & 367  \\
1.28(9)  & 0.8 & 1.032 & \ \ 4000 &  524  \\
1 & 0.0 & 1.605 &         \ \ 2000 & 248  \\
1 & 0.0 & 1.669 &         \ \ 4000 & 575  \\
1 & 0.0 & 1.7024 &     \ \  8000 & 489  \\
1 & 0.0 & 1.7325 &     \ \ 16000 & 501  \\
1 & 0.0 & 1.75665 &     \ \ 32000 & 1218  \\
0.80(4) & $-0.6$ & 2.45 & \ \ 4000 & 414  \\
0.70(4) & $-0.8$ & 3.0 &  \ \ 8000 & 1486  \\
0.70(4) & $-0.776$ & 3.0 & \ \ 16000 & 2341 \\
\hline
\end{tabular}
\caption{The parameters of the ensembles used in this work.  The first column shows the relative lattice spacing, with the ensembles at $\beta=0$ serving as the fiducial lattice spacing.  The quoted error is a systematic error associated with finite-volume effects.  The second column is the value of $\beta$, the third is the value of $\kappa_2$, the fourth is the number of four-simplices in the simulation, and the fifth is the number of configurations sampled.}
\label{tab:ensembles}
\end{center}
\end{table}

\section{Theoretical Background}
\label{sec:theo-back}
In this section we give the necessary theoretical background for extracting and interpreting the numerical data associated with the binding energy of two scalar particles.  We begin with a discussion of scalar fields on dynamical triangulations, then we show how to calculate the binding energy between scalar particles on our lattices, and finally we review the nonrelativistic limit of the binding energy in the continuum limit.

\subsection{Scalar fields on dynamical triangulations}
\label{subsec:scalar-dts}
In order to map Eq.~\eqref{eq:matter-action} on to the lattice, we use a naive discretization where we associate each scalar field at $x$, $\phi(x)$, with a four-simplex (or equivalently a dual-lattice site).  In this way each scalar field is restricted to only have five neighbors through the five surrounding tetrahedra (or the dual edges).  The lattice action on the dual lattice can then be written as
\begin{align}
\label{eq:lat-scalar}
    S_{M}^{\text{lat}} = \frac{1}{2}\sum_{\langle x y \rangle} (\phi_{x} - \phi_{y})^{2} + \frac{m_{0}^{2}}{2} \sum_{x} \phi_{x}^{2}
\end{align}
where $\langle x y \rangle$ is a nearest-neighbor pair between two dual lattice sites (or two four-simplices) which is counted once, and $m_{0}$ is the bare mass.  Note that the lattice scalar action has a shift symmetry in the absence of a mass term \cite{Jha:2018xjh,Catterall:2018dns,Catterall:2018lkj}.  This is also true of the continuum action Eq.~(\ref{eq:matter-action}) for zero mass and in the absence of any additional scalar self interaction terms.  This shift symmetry ensures that the renormalized mass goes to zero as the bare mass approaches zero without any fine tuning.  It is a non-trivial check of the calculations that this is the case.

Expanding and collecting terms above---as well as noting that we work on a compact topology---and normalizing the coefficient of the kinetic term to one, we can write Eq.~\eqref{eq:lat-scalar} as
\begin{align}
\label{eq:scalar-lap}
    S_{M}^{\text{lat}} = \sum_{x,y} \phi_{x} L_{xy} \phi_{y}
\end{align}
with $L$ the dual lattice Laplacian plus a mass term, and the sums over $x$ and $y$ are unrestricted over the whole lattice.  In the continuum $L$ is the Klein-Gordon operator and the inverse of the $L$ matrix is the scalar propagator.  In the next subsection we review the form of the binding energy between scalar particles in the presence of gravity on the lattice.

\subsection{The binding energy}
\label{subsec:binding-energy}
The binding energy between two particles is given by $E_{b} = 2m - M$, where $m$ is the mass of a single particle, and $M$ is the mass of the two-body bound state.  If $E_{b}$ is greater than zero, it indicates that the two-body state has a lower energy than twice the energy of the single-particle state, and thus a bound state can form.

To compute the binding energy between two scalar particles, we must first calculate the propagator for a free scalar field living on a triangulation.   
Given the lattice scalar action of Eq.~\eqref{eq:scalar-lap}, we need only invert the matrix $L_{xy}$ to obtain the propagator.
For an arbitrary triangulation, we can write down $L_{xy}$ explicitly,
\begin{equation}
    L_{xy} = (D_{x}+m_{0}^{2})\delta_{xy} - A_{xy}
\end{equation}
where $D_{x}$ is the number of neighboring four-simplices to a four-simplex (in our case always five), $m_{0}$ is the bare input mass, $\delta_{xy}$ is the Kronecker delta, and $A_{xy}$ is the adjacency matrix, which has matrix elements
\begin{equation}
    A_{xy} = 
    \begin{cases}
        1 & \text{if $x$ and $y$ share a dual edge} \\
        0 & \text{otherwise.}
    \end{cases}
\end{equation}
The matrix elements of $L^{-1}$ contain the propagators between two four-simplices.  To compute the binding energy, we need the squares of the propagators as well, $(L_{xy}^{-1})^{2}$.  The interpretation of $(L^{-1}_{xy})^{2}$ is the two-particle propagator between $x$ and $y$.

With the one- and two-particle propagators, we can compute the average two-point correlation function as a function of geodesic distance on the lattice.  The one- and two-particle, two-point correlation functions are, respectively \cite{deBakker:1996qf},
\begin{equation}
\label{eq:1p-corr}
    G(r) = \left\langle \frac{\sum_{x,y} L_{xy}^{-1} \delta_{|x-y|,r}}{\sum_{x,y} \delta_{|x-y|,r}} \right\rangle
\end{equation}
and
\begin{equation}
\label{eq:2p-corr}
    G^{(2)}(r) = \left\langle \frac{\sum_{x, y} (L_{xy}^{-1})^{2} \delta_{|x-y|,r}}{\sum_{x, y} \delta_{|x-y|,r}} \right\rangle
\end{equation}
as a function of $r$, the geodesic distance between dual lattice points.  In Eqs.~\eqref{eq:1p-corr} and~\eqref{eq:2p-corr} there are two averages shown.  The first is the average over all possible distances a separation $|x-y| = r$ away.  In practice this is not done in its entirety, and a fixed number of source points are chosen, from which propagators and distances are calculated to all other points.  The second average is indicated with the angled brackets, which denote an ensemble average---an average over configurations.

From $G$ and $G^{(2)}$ it is possible to compute the binding energy between two scalar masses.  We can use the asymptotic forms of $G$ and $G^{(2)}$, which are expected to feature an exponential fall-off for large distances,
\begin{equation}
\label{eq:correlators}
    G(r) \propto \frac{e^{-m r}}{r^p}, \quad G^{(2)} \propto \frac{e^{-M r}}{r^{q}},
\end{equation}
to extract the binding energy.  Above, $m$ is the renormalized one-particle mass, and $M$ is the two-particle renormalized mass.  Consider the function,
\begin{equation}
\label{eq:f-func}
    F(r) = \frac{G^{(2)}}{G^{2}},
\end{equation}
which for sufficiently large source-sink separations becomes
\begin{align}
    F(r) \propto \frac{e^{-(M - 2m) r}}{r^{q-2p}}.
\end{align}
We can identify the binding energy with the exponent, 
\begin{align}
\label{eq:binding}
    E_{b} \equiv 2m - M   ,
\end{align}
and we can identify the power-law exponent, $\gamma \equiv q - 2p$.  Thus $E_{b}$ measures the difference in energy between a two-body system in its ground state and twice the mass of a single particle in its ground state, with gravitational effects taken into account.  

With these definitions, taking the logarithm of $F$, we find
\begin{equation}
\label{eq:log-F}
    \log F(r) \simeq E_{b} r + Z_{E} - \gamma \log r ,
\end{equation}
where $Z_{E}$ is a constant.  The single particle propagator has the same form after taking the logarithm,
\begin{equation}
\label{eq:log-G}
    \log G(r) \simeq -m r + Z_{G} - p \log r.
\end{equation}
Besides the constant shifts present in both equations, the renormalized mass and the binding energy both appear as linear terms, while the power-law behavior appears as a logarithmic correction.  In the physically relevant case, and using our conventions, $E_{b}$ and $m$ should be positive, indicating an attractive gravitational force and particles with positive energy density, respectively.  In the next subsection we consider the non-relativistic limit of the binding energy between two scalar masses.

\subsection{Binding energy in the non-relativistic limit}
\label{sec:newton}
In this subsection we consider the energy levels associated with gravitational bound states. We follow Ref.~\cite{deBakker:1996qf} in assuming that the mass of the constituent particles is much lighter than the Planck mass such that the particles are weakly coupled.  In this case the magnitude of the binding energy is well below the mass of the constituent particles and we can work in the nonrelativistic limit.
In that limit the effective description for the bound states of two scalar masses by gravity is simply given by Schr\"{o}dinger's equation with Newton's gravitational potential.  The bound state solution is then just the Schr\"{o}dinger solution for positronium, but with the fundamental constants that appear in the Coulomb potential exchanged for those of Newton's potential.  

Thus we consider the Schr\"{o}dinger equation,
\begin{align}
    -\nabla^{2} \psi(r,\theta,\phi) + 2\mu(U(r)-E) \psi(r,\theta,\phi) = 0,
\end{align}
with potential
\begin{align}
    U(r) = -\frac{G m^2}{r},
\end{align}
where $m$ is the particle mass, $\mu$ is the reduced mass equal to $m/2$ in the degenerate case considered here, and $G$ is Newton's constant.  The solution to this equation for a potential of this form is well known.
The energy eigenvalues, given by the principal quantum number $n$, are
\begin{align}
\label{eq:newt-binding}
    E_{n} = \frac{G^2 m^5}{4 n^2}
\end{align}
in units where $\hbar = c = 1$.  

The ground state binding energy for two identical scalar masses held together by gravity is therefore $E_{1} = m^5 G^2 / 4$, giving us a clear prediction for the dependence of the binding energy on the constituent mass.  However, in order to make contact with this power law using lattice methods we first need to take the continuum, infinite-volume limit of our calculation.  To get an idea for what type of discretization and finite-volume effects we might expect, we recall that the fractal dimension of the geometries varies as a function of distance scale.  Even at the largest distance scales that we can probe, the spectral dimension measured on the lattices does not get much above three.  It is only in the continuum, infinite-volume limit that the spectral dimension at long distances extrapolates to four, as it must to reproduce our world.  In a 2+1 dimensional world, the same derivation for the binding energy would lead to $E_1 \propto m^2$, so we might expect a large extrapolation of the exponent in the power law, from $\sim 2$ to $\sim 5$, if the scalar fields on our lattices see a long distance effective dimension of around three at coarse couplings and small volumes.  This expectation is born out by the data, as we show in the following section.  

To further interpret our results, it is useful to model the binding energy as a function of the dimension $d$.  By dimensional analysis, we have $E_1 \propto m$ in 1+1 dimensions.  Taking a simple quadratic fit to the three values of the exponent quoted here for integer dimension leads to the scaling $E_1 \propto m^\alpha$, with
\bea \label{eq:alpha} \alpha= d^2 -4d +5.
\eea
The values of $d$ that determine this dependence interpolate over the range of effective long distance dimensions that were found for our lattices in Ref.~\cite{Laiho:2016nlp}, so this simple estimate of the relationship between $\alpha$ and $d$ should be reliable enough to determine what numerical value of $d$ is implied by the $\alpha$ obtained from our simulations.

The other limit that must be taken in our calculation is the non-relativistic limit.  Again following Ref.~\cite{deBakker:1996qf}, we can get a rough estimate of the size of relativistic corrections to Eq.~(\ref{eq:newt-binding}).  Consider the Hamiltonian
\begin{align}
\label{eq:hamiltonian-binding}
    H = 2\sqrt{m^2+p^2}-\frac{G m^2}{r}.
\end{align}
If we replace the momentum $p$ by $1/r$ and minimize the energy, we find
\begin{align}
\label{eq:relativistic-binding}
    E_b = 2m - 2m\sqrt{1-\frac{G^2 m^4}{4}}.
\end{align}
This implies that $G m^2/2~\ll~1$ in the non-relativistic limit.  The region where we see power law behavior for the binding energy as a function of scalar mass in our numerical data is compatible with this condition.

    

\section{Numerical results}
\label{num-res}
In this section we discuss the details of the numerical analysis of the correlators, which yields the binding energy and renormalized mass.  We then explain the details of the power-law fits to the binding energy versus the renormalized mass.  Finally, we discuss the infinite-volume, continuum extrapolation of the exponent in the power law as well as our determination of Newton's constant.

\subsection{Correlation functions}
\label{subsec:corr-funcs}
The correlators defined in Eqs.~\eqref{eq:1p-corr} and~\eqref{eq:2p-corr} are obtained from exact inversions of the matrix $L_{xy}$ calculated on a given lattice configuration for a given bare mass value.  We do not consider every simplex on the lattice as a possible source.  Instead, we vary the number of sources on each configuration from one, five, 20, and 60 in order to assess the effect the number of source simplices used has on the statistical error.  We find that for a large number of sources, say 60, the systematic errors associated with modeling the deviations of the data from the model fit function are dominant.  These deviations could be due to excited states, and finite-size and discretization effects.  In order to avoid this difficulty of modeling the systematics, we use a single source per configuration in our main analysis.  All source simplices are selected randomly from the largest three-volume cross-section of the entire lattice.  This is done by first shelling a configuration starting from a source chosen at random, and then only selecting sources for the propagator from the largest slice in the shelling.  We find that restricting our sources to come from the largest three-slice minimizes finite lattice spacing effects, and it is the same procedure that we have used in previous work on the spectral dimension \cite{Laiho:2016nlp} and for our studies of K\"{a}hler-Dirac fermions \cite{Catterall:2018dns}.

\begin{figure}
    \centering
    \includegraphics[width=8.6cm]{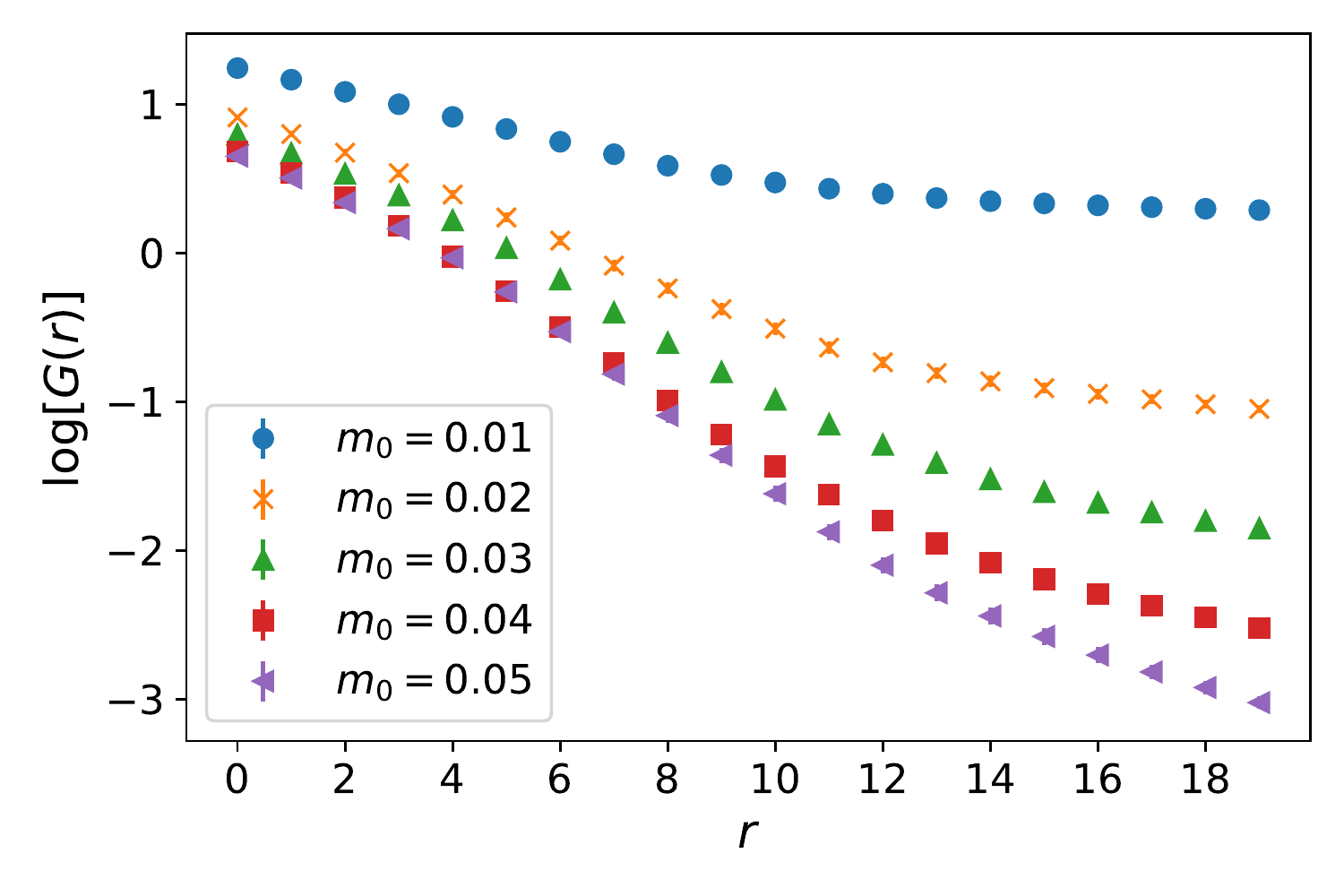}
    \caption{The logarithm of the two-point correlator, $G(r)$, for five masses on the $N_{4} = 8000$, $\beta = 0$ ensemble.  We see a bend in the data which is pushed progressively out to larger $r$ values as the bare mass is increased.  The distances displayed on the horizontal axis are in units of the distance between the centers of adjacent four-simplices, \emph{i.e.} a dual edge length.}
    \label{fig:corr-example}
\end{figure}

\begin{figure}
    \centering
    \includegraphics[width=8.6cm]{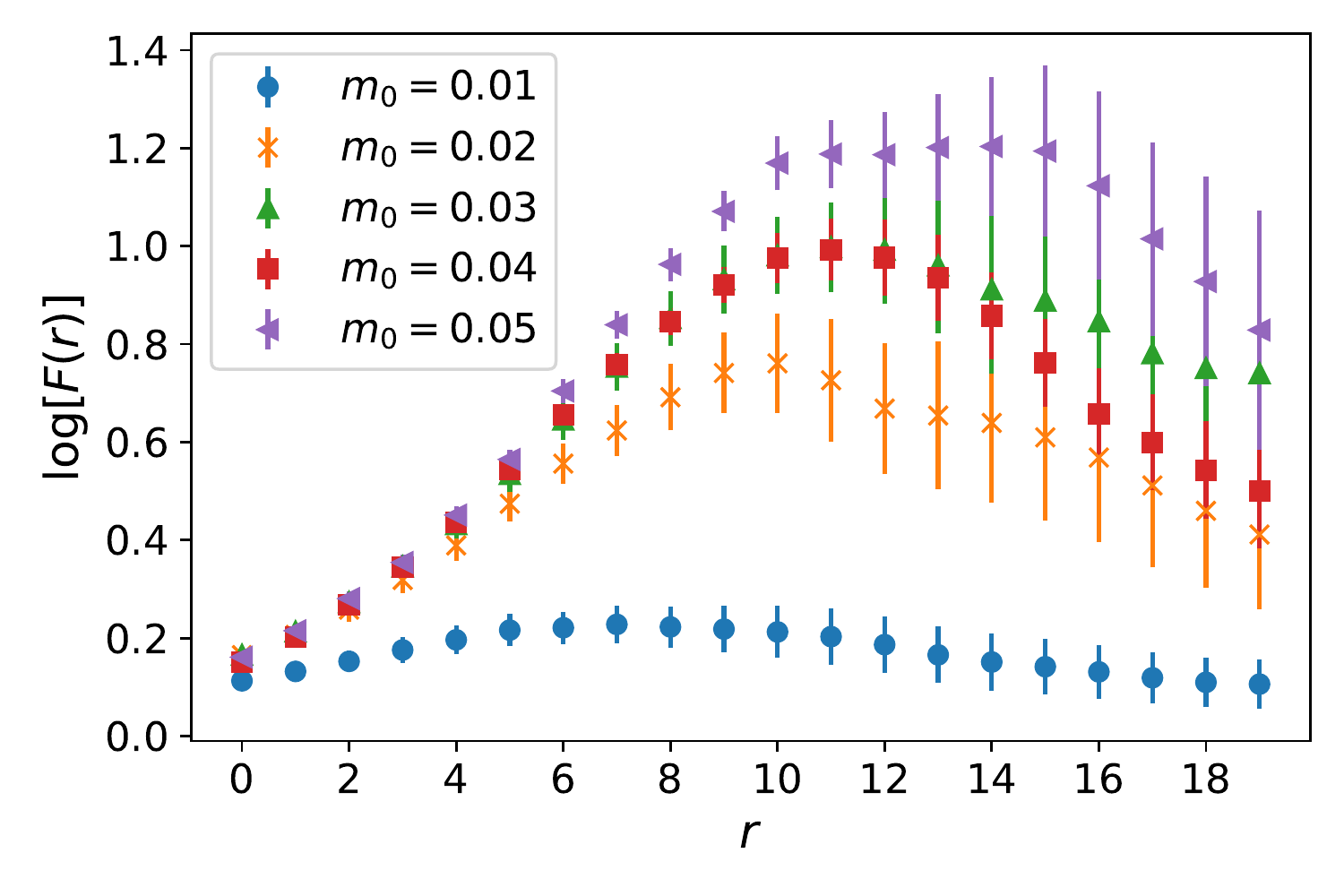}
    \caption{The logarithm of the ratio between the two-particle, two-point correlator, and the square of the one-particle, two-point correlator, $F(r)$.  Five different bare masses are shown on the $N_{4} = 8000$, $\beta = 0$ ensemble.  We see a peak in the data, separating a positively sloped region and a negatively sloped region, which is pushed to larger $r$ values for larger bare mass values.  The distances displayed on the horizontal axis are in units of the distance between the centers of adjacent four-simplices, \emph{i.e.} a dual edge length.}
    \label{fig:f-example}
\end{figure}

\begin{figure}
    \centering
    \includegraphics[width=8.6cm]{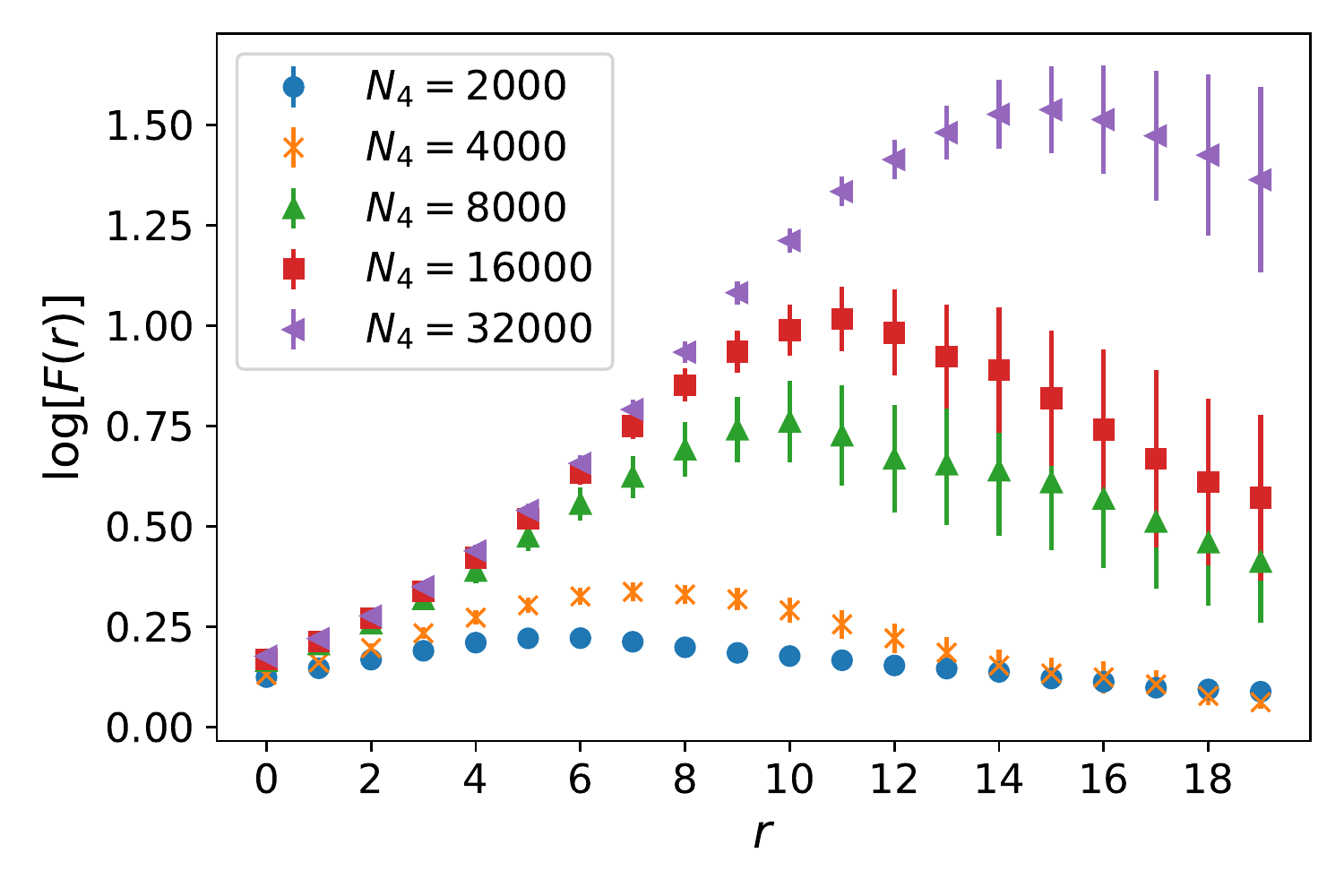}
    \caption{The logarithm of the ratio of the two-particle correlator to the square of the one-particle correlator as a function of distance for multiple volumes at $\beta = 0$, and $m_{0} = 0.02$.  We see as the volume is increased the peak is pushed to larger values of $r$ indicating the turn-over in the data is most likely a long-distance lattice artifact.  The distances displayed on the horizontal axis are in units of the distance between the centers of adjacent four-simplices, \emph{i.e.} a dual edge length.}
    \label{fig:volvary}
\end{figure}

An example of correlator data is shown in Fig.~\ref{fig:corr-example}, and an example of the ratio $F$ defined in Eq.~\eqref{eq:f-func} is shown in Fig.~\ref{fig:f-example}.  Both plots use log-linear coordinates and show results for several masses.
These figures display a feature that appears across all ensembles, to varying degrees.  We see around $r \approx 10$ lattice spacings a bend in the correlator, and a peak where the derivative of $\log[F]$ changes sign.  We also see this feature is a function of the bare mass, and as the bare mass is increased, this bend is pushed out further to larger distances.  The same thing happens as the volume of the system is increased.  In Fig.~\ref{fig:volvary}
we see the peak is pushed to larger distances as the system volume is increased.  This indicates that the turn-over in the data is most likely due to long-distance lattice artifacts.  As noted in Ref.~\cite{Laiho:2016nlp}, there are baby universes that branch off of the mother universe, where the baby universes can be quite long, although their cross-section is of order the lattice spacing.  This effect is most pronounced on our coarsest lattices, where it can significantly modify long-distance physics, although the effect appears to vanish in the continuum limit.  It is useful to keep this in mind when choosing a fit window to extract masses from our correlation functions, since it sets an upper bound on how far in the Euclidean time extent we can fit and still expect our model fit function to describe the data.  This bend in the data that we see is most likely due to baby-universe effects at long distances.  At short distance scales we expect the usual discretization effects, as well as excited state contamination.  Thus, the fit window to the correlation function is rather constrained in our current approach.

The peak in $\log[F]$ is one of the first clues on how to extract a physically motivated answer for the binding energy.  From the definition of $E_{b}$ in Eq.~\eqref{eq:binding} the coefficient of $r$ should be positive if the two-particle state is bound.  We see this is only possible between $r=0$ and the peak around $r \approx 10$ lattice spacings (for the specific ensemble in Figs.~\ref{fig:corr-example} and~\ref{fig:f-example}).  In fact, the existence of such a region is already encouraging, since it implies there exists an attractive force between scalar masses inside the dynamical triangulations framework.  This was noticed already in Ref.~\cite{deBakker:1996qf}.

Additionally, looking at Fig.~\ref{fig:f-example} we can see a change in concavity for the larger masses around a value of $r \approx 5$.  This inflection point marks the change of the concavity from a region that is concave up, to a region where the data turns over \emph{i.e.} the peak.  This inflection point denotes the end of the valid fitting region according to Eq.~\eqref{eq:log-F}, since after this point the long-distance effects begin to dominate the shape of the function.  Across all bare masses and ensembles, we fit to a region that begins at $r=1$ and ends around the inflection point of $\log[F]$.  We fit this same range in the one-particle correlator, and the $F$ function.

The choice of fit function is decided by Eqs.~\eqref{eq:log-F} and~\eqref{eq:log-G}.  Thus, we use a function of the form,
\begin{align}
\label{eq:one-fit}
    f(r) = X r + Y + Z \log r
\end{align}
for both the $\log[F]$ and $\log[G]$ data, with $X$, $Y$, and $Z$ as fit parameters.
By fitting the $\log[F]$ and $\log[G]$ data to the functional form in Eq.~\eqref{eq:one-fit} we can extract the binding energy, the renormalized mass, and the exponents $\beta$ and $\gamma$ as a function of the bare mass.

\begin{figure}
    \centering
    \includegraphics[width=8.6cm]{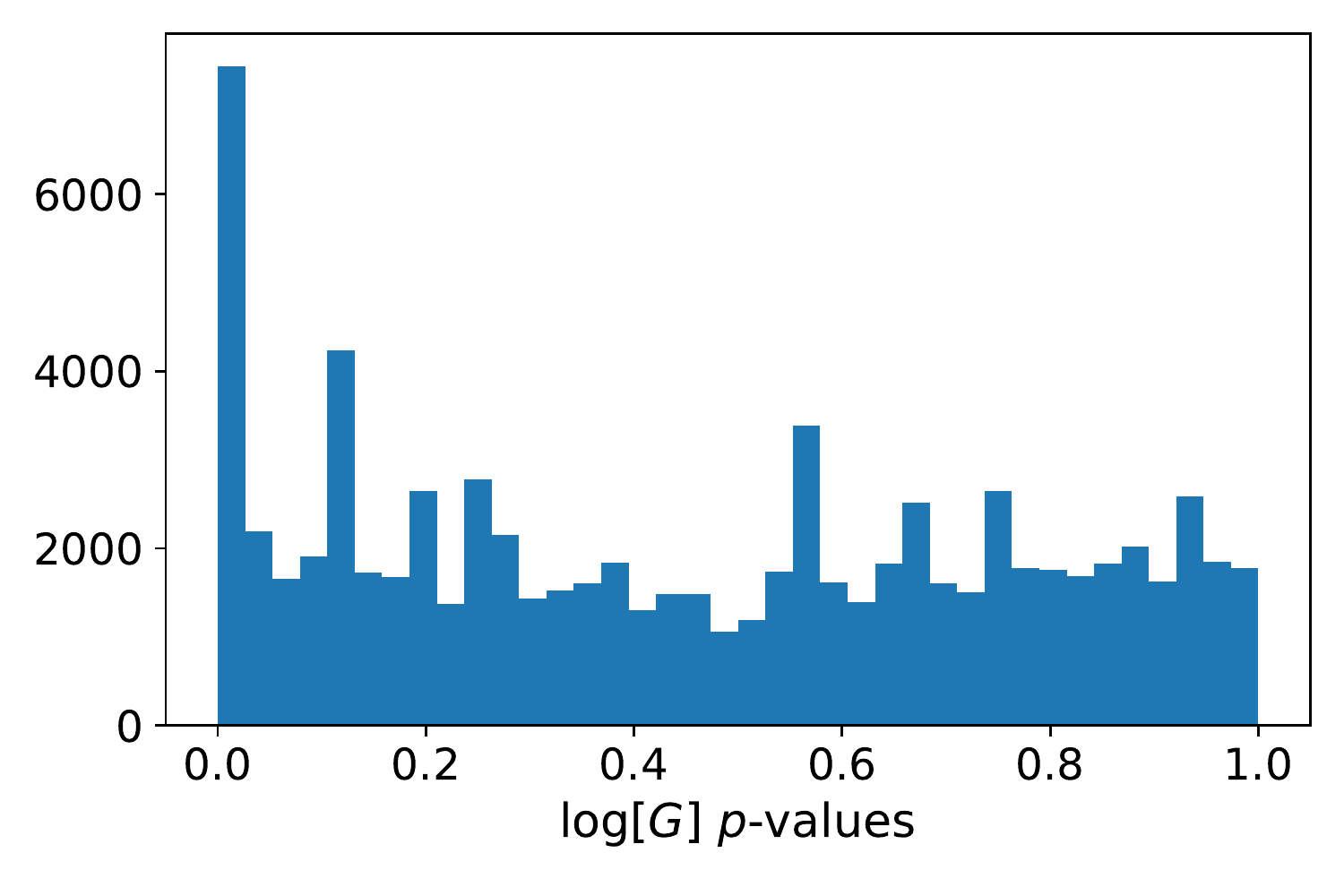}
    \caption{A histogram of the $p$-values extracted from fits to $\log[G]$.  This histogram contains the $p$-values from the individual jackknife fits for all ensembles and mass values used downstream in the analysis, combined into a single data set.}
    \label{fig:pv-G}
\end{figure}

\begin{figure}
    \centering
    \includegraphics[width=8.6cm]{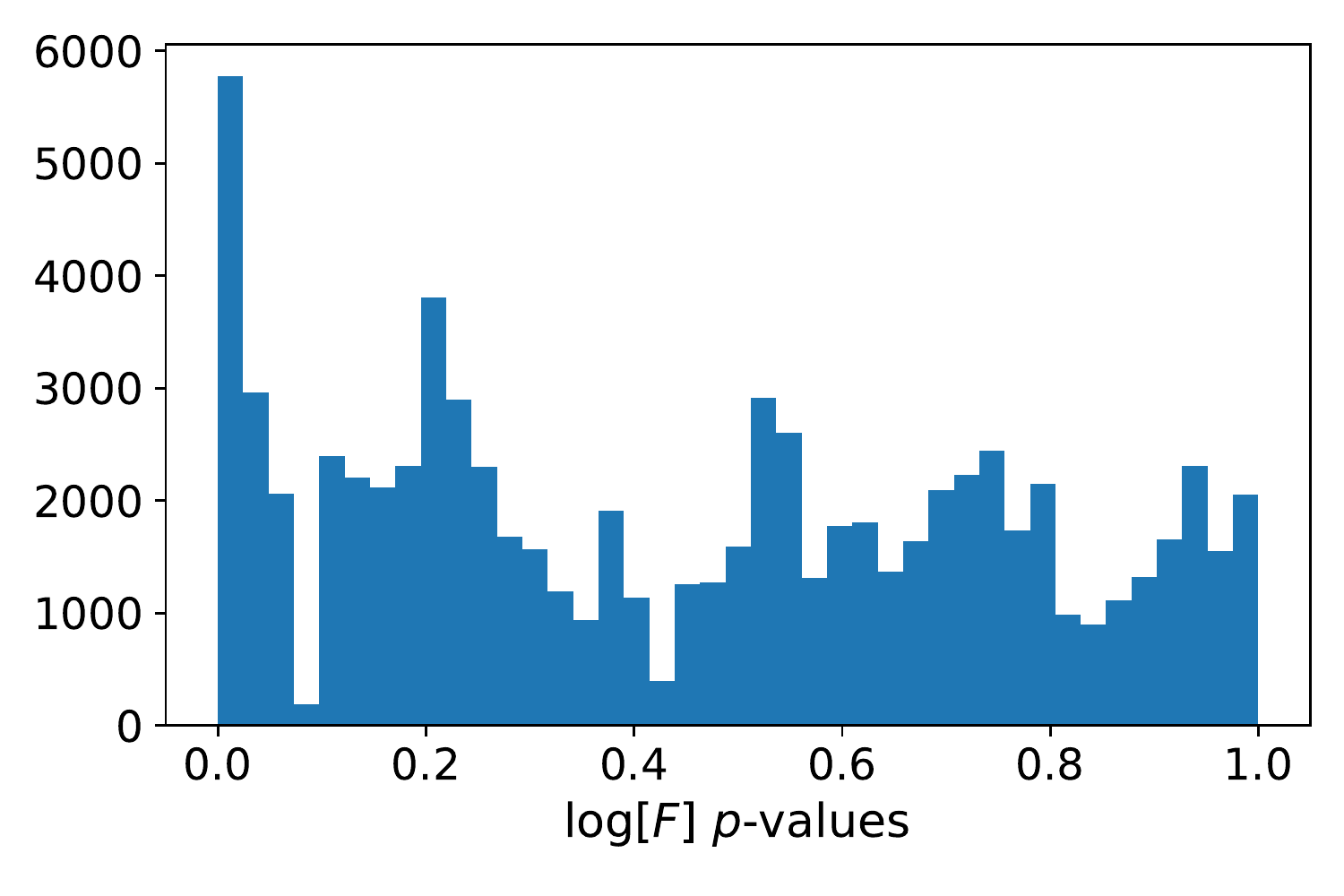}
    \caption{A histogram of the $p$-values extracted from fits to $\log[F]$.  This histogram contains the $p$-values from the individual jackknife fits for all ensembles and mass values used downstream in the analysis, combined into a single data set.}
    \label{fig:pv-F}
\end{figure}

\begin{figure}
    \centering
    \includegraphics[width=8.6cm]{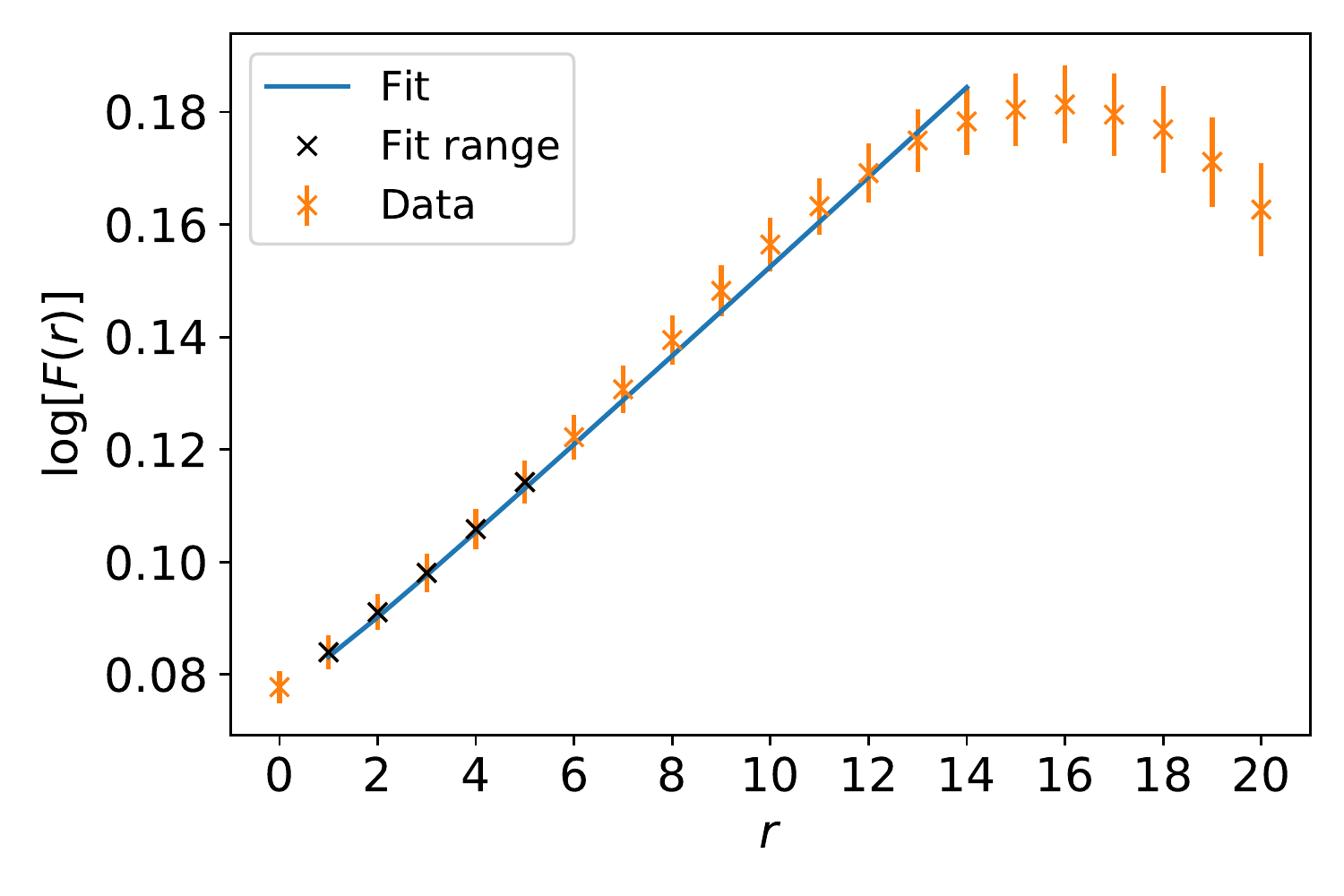}
    \caption{An example of a fit to the $N_{4} = 16,000$ simplex ensemble with $\beta = -0.776$ for $\log[F]$ at bare mass $m_{0} = 0.004$.  Here the chosen fit range is shown in black, along with the best fit line. The $\chi^{2}/\text{d.o.f} = 0.77$ for the fit at this mass and this ensemble corresponding to a $p$-value of $0.46$.}
    \label{fig:logF-fit}
\end{figure}

\begin{figure}
    \centering
    \includegraphics[width=8.6cm]{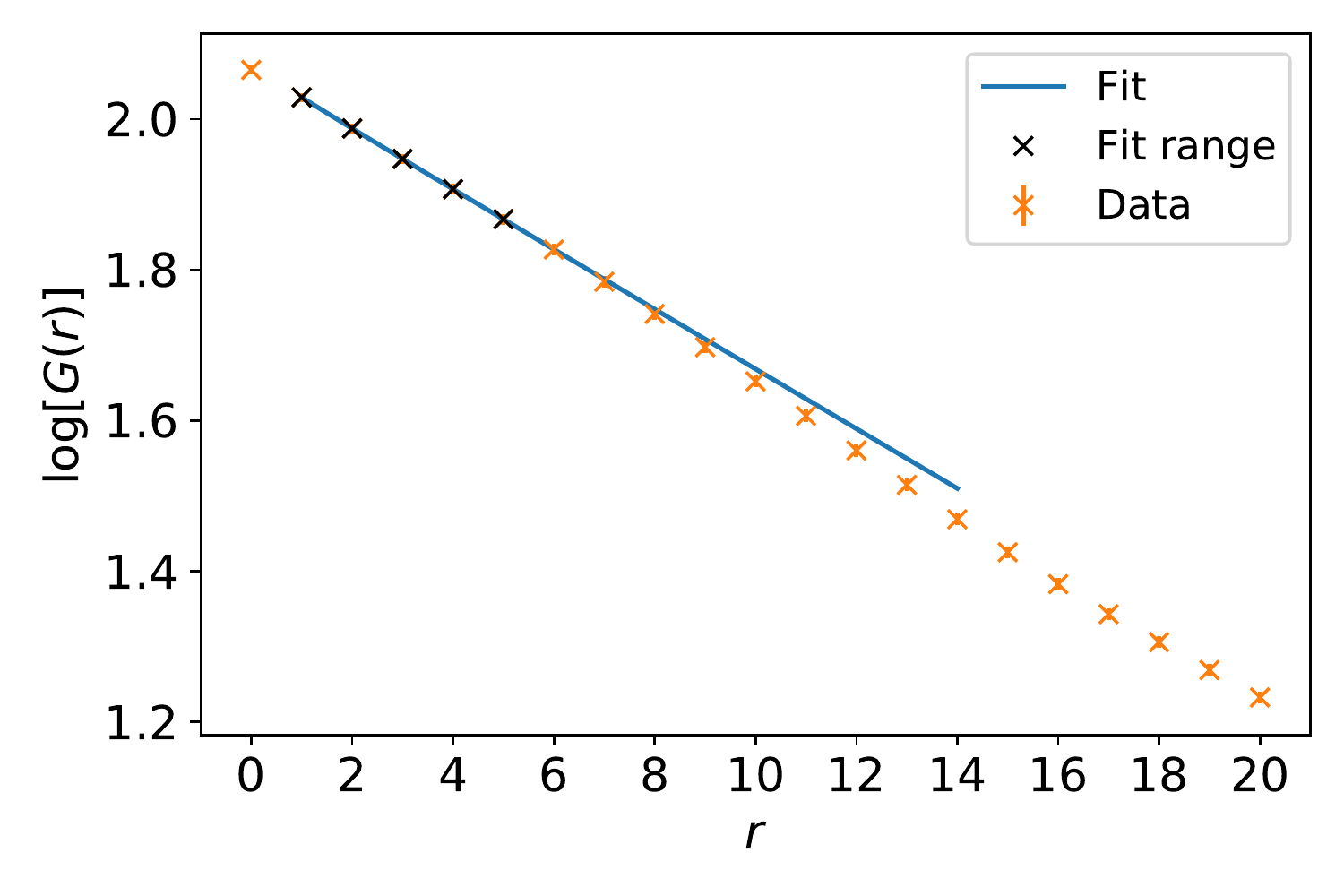}
    \caption{An example of a fit to the $N_{4} = 16,000$ simplex ensemble with $\beta = -0.776$ for $\log[G]$ at bare mass $m_{0} = 0.004$.  Here the chosen fit range is shown in black, along with the best fit line.  The $\chi^{2}/\text{d.o.f} = 0.19$ for the fit at this mass and this ensemble corresponding to a $p$-value of $0.83$.}
    \label{fig:logG-fit}
\end{figure}

The fits are done with non-linear least squares fitting including the correlations of the dependent data.  The errors are estimated using single-elimination jackknife resampling, including the off-diagonal terms in the correlation matrix.  The size of autocorrelation errors is estimated using a blocking procedure; the data is blocked until the errors no longer increase.  In order to retain enough information to resolve the correlation matrix when performing fits, the data is not blocked, but the errors are inflated to reflect the increased error due to autocorrelations.  The fits are performed under a jackknife, and the correlation matrix is reconstructed for each individual fit under the jackknife from the data on each jackknife sub-ensemble.  By including correlations in the fit, the $\chi^2$ per degree of freedom is expected to be a reliable measure of goodness of fit.  We compute from the $\chi^2$ and the number of degrees of freedom a confidence interval (a $p$-value) for the fit, correcting for finite sample size.  We make a histogram of $p$-values from the fits for which the fit parameters are propagated through to the rest of the analysis.  This includes fits from all ensembles.  The resulting histogram of $p$-values is relatively uniform, and is shown in Figs.~\ref{fig:pv-G}, and~\ref{fig:pv-F} for the correlator fits, and $F$ fits, respectively.  
Only the lowest bin possesses a small spike.  Since the fits in this bin are scattered throughout the parameter values of the analysis more or less at random, we do not ascribe an additional error to this slight deviation from a flat distribution.
An example of the fit for the $N_{4} = 16,000$ simplex ensemble with $\beta = -0.776$ to the $\log[F]$ and $\log[G]$ data can be seen in Figs.~\ref{fig:logF-fit} and~\ref{fig:logG-fit}.  Given the results for the binding energy and the renormalized mass for a wide range of bare mass values on many different ensembles, we are able to test the theory presented in Sect.~\ref{sec:newton}.  This is done in the following subsections.

\subsection{Mass dependence of the binding energy}
\label{subsec:mass-be}
\begin{figure}
    \centering
    \includegraphics[width=8.6cm]{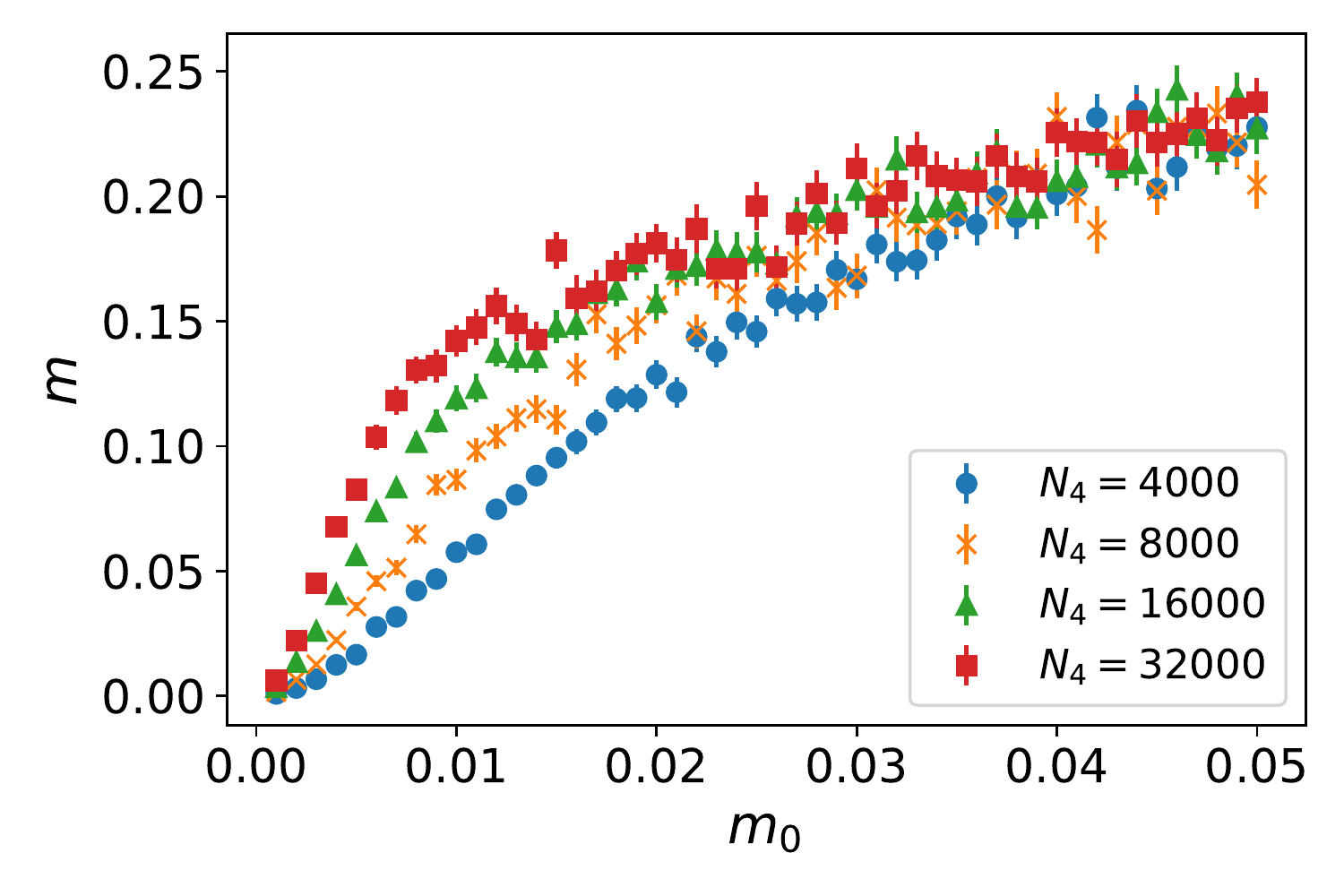}
    \caption{The renormalized mass plotted against the bare mass for four volumes with $\beta = 0$.  Here we see the renormalized mass is multiplicatively renormalized for sufficiently small bare masses.}
    \label{fig:shift}
\end{figure}

The dependence of the renormalized mass on bare mass is shown in Fig.~\ref{fig:shift} for four different volumes at fixed lattice spacing ($\beta=0$).  It is clear from this plot that the renormalized mass goes to zero as the bare mass also approaches zero, which is a consequence of the shift symmetry of the lattice action. This provides a useful check of our calculation.

\begin{figure}
    \centering
    \includegraphics[width=8.6cm]{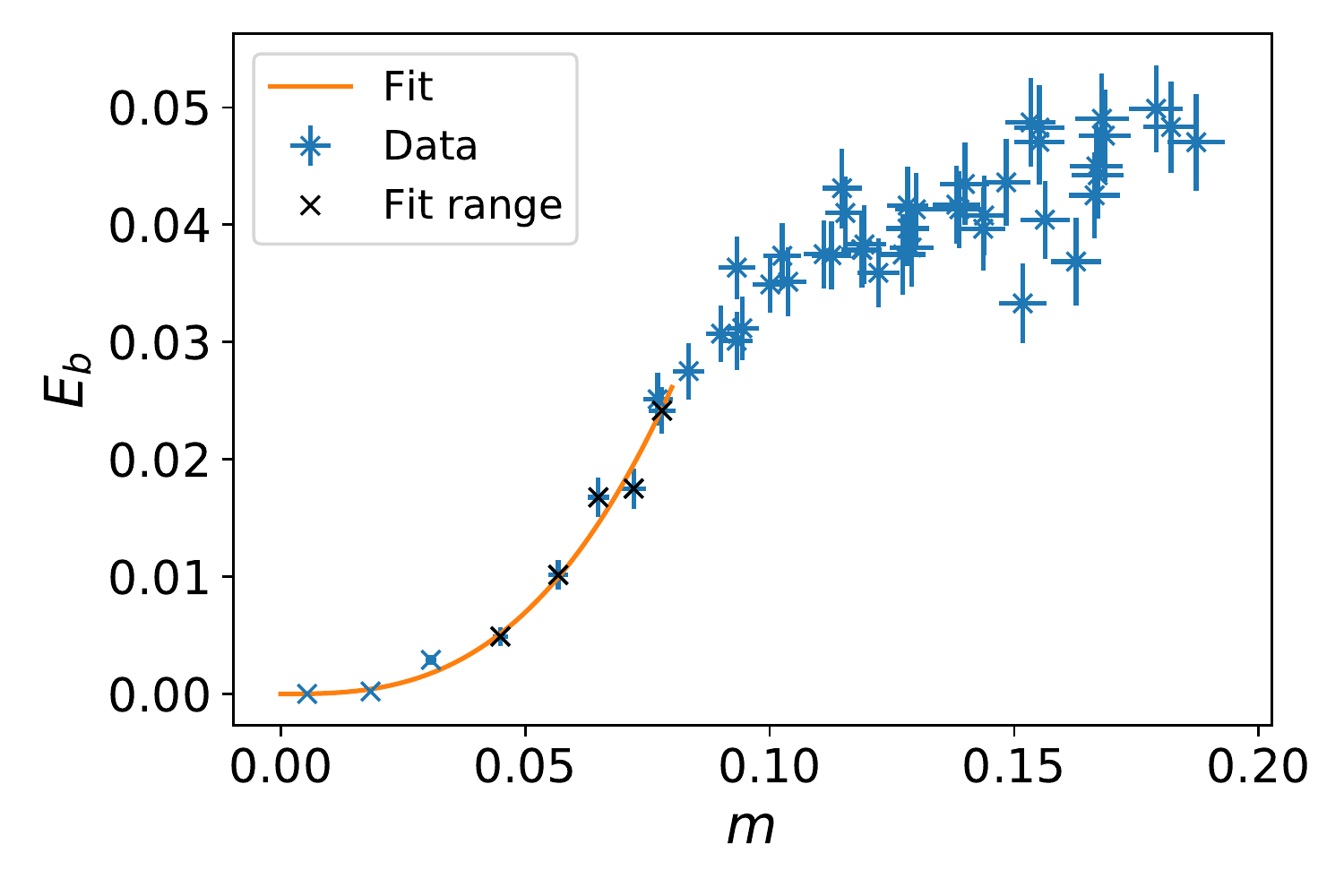}
    \caption{The power-law fit to the binding energy plotted against the renormalized mass for the $N_{4} = 16,000$, $\beta = -0.776$ ensemble.  The fit range is shown in black, and the solid line is the fit to the data.  The fit corresponds to a $\chi^{2}$/d.o.f. $= 0.59$, with a $p$-value of 0.62.}
    \label{fig:be-rm_16kb-0p776}
\end{figure}

\begin{figure}
    \centering
    \includegraphics[width=8.6cm]{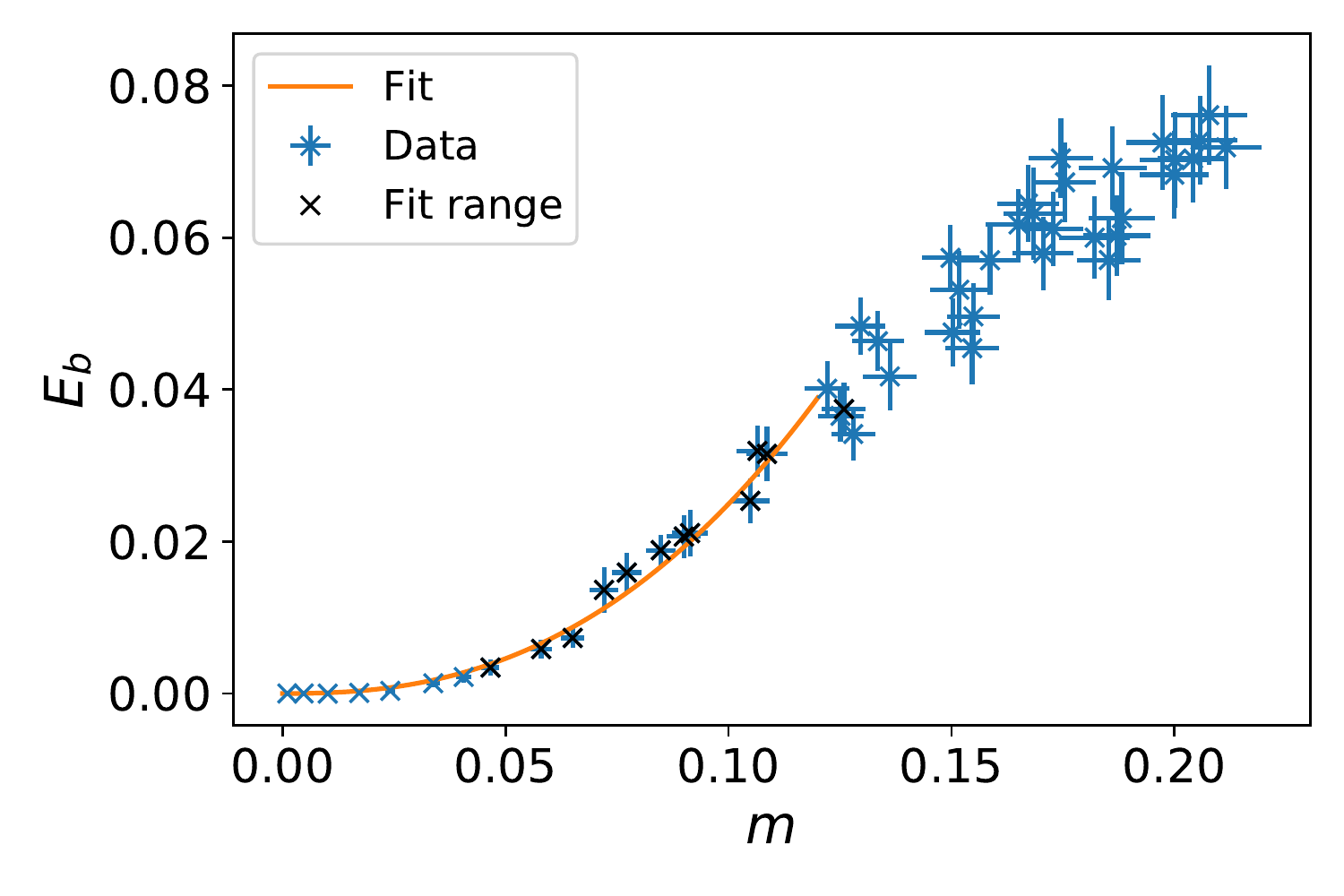}
    \caption{The power-law fit to the binding energy plotted against the renormalized mass for the $N_{4} = 4,000$, $\beta = -0.6$ ensemble.  The fit range is shown in black, and the solid line is the fit to the data.  The fit corresponds to a $\chi^{2}$/d.o.f. $= 0.64$, with a $p$-value of 0.79.}
    \label{fig:be-rm_4kb-0p6}
\end{figure}

\begin{figure}
    \centering
    \includegraphics[width=8.6cm]{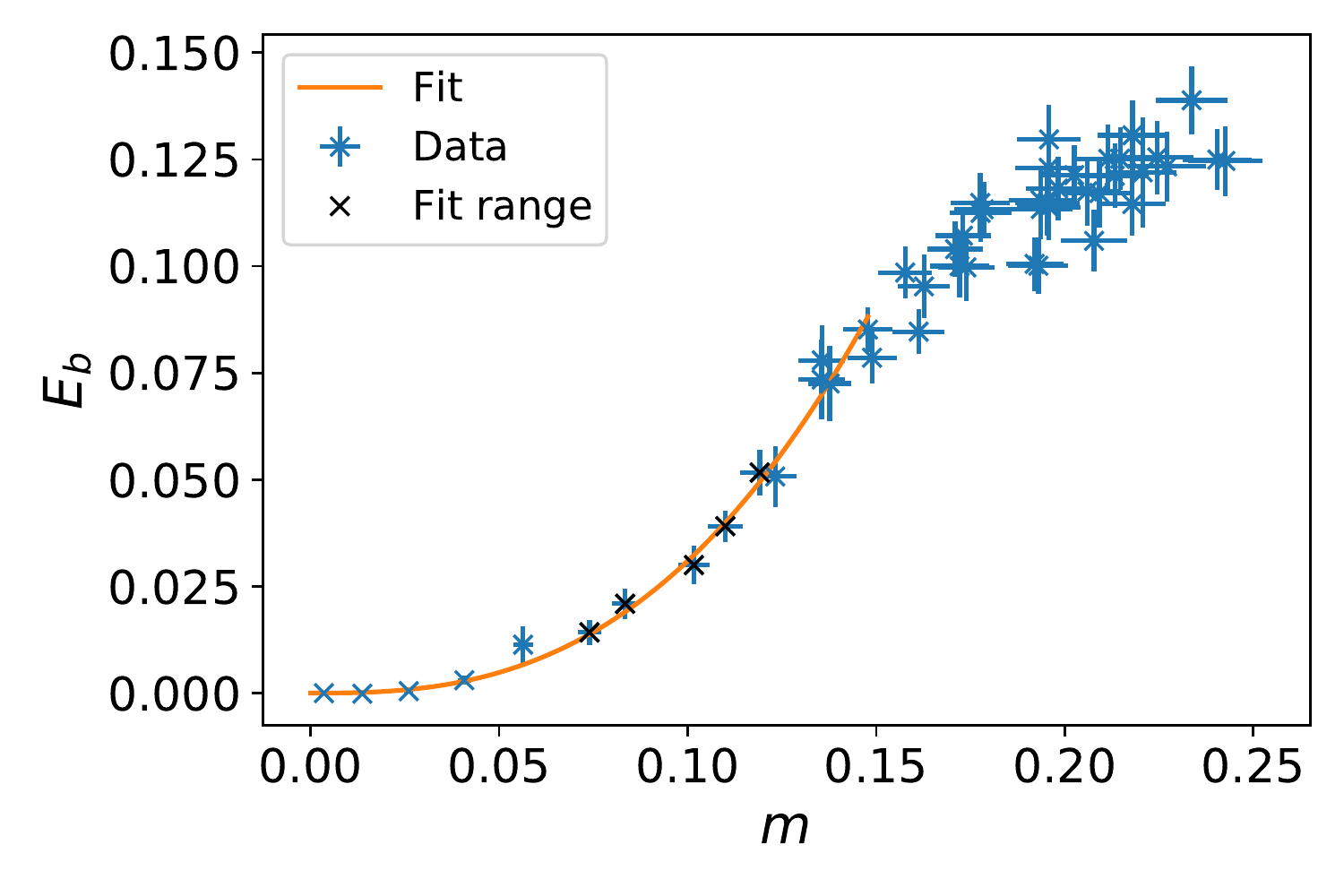}
    \caption{The power-law fit to the binding energy plotted against the renormalized mass for the $N_{4} = 16,000$, $\beta = 0$ ensemble.  The fit range is shown in black, and the solid line is the fit to the data.  The fit corresponds to a $\chi^{2}$/d.o.f. $= 0.15$, with a $p$-value of $0.93$.}
    \label{fig:be-rm_16kb0}
\end{figure}

\begin{figure}
    \centering
    \includegraphics[width=8.6cm]{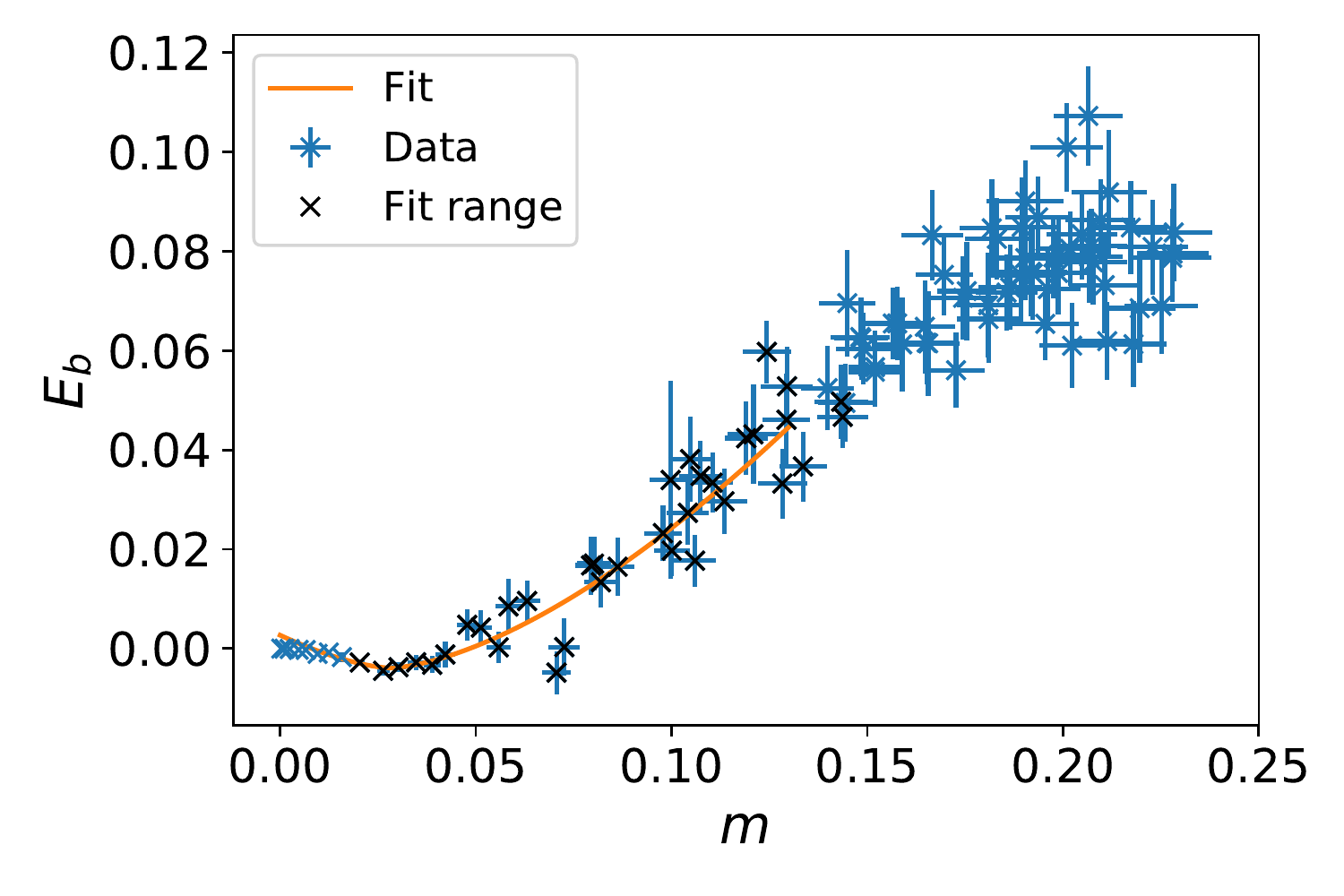}
    \caption{The power-law fit to the binding energy plotted against the renormalized mass for the $N_{4} = 4,000$, $\beta = 1.5$ ensemble.  The fit range is shown in black, and the solid line is the fit to the data.  The fit corresponds to a $\chi^{2}$/d.o.f. $= 1.16$, with a $p$-value of $0.31$.}
    \label{fig:be-rm_4kb1p5}
\end{figure}

The dependence of the binding energy on the renormalized mass is shown in Figs.~\ref{fig:be-rm_16kb-0p776}, \ref{fig:be-rm_4kb-0p6}, \ref{fig:be-rm_16kb0} and~\ref{fig:be-rm_4kb1p5} for four different ensembles.
In order to make contact with Newtonian gravity, we must look for a power-law dependence for the binding energy as a function of renormalized mass, as given in Eq.~(\ref{eq:newt-binding}).  As a first step, in order to eventually be able to compare results across lattice spacings, we put the results in the same lattice spacing units.  Before performing the fits we re-scale all the binding energies and renormalized masses to that of the fiducial lattice spacing at $\beta=0$ using the relative lattice spacings given in table~\ref{tab:ensembles}.

To study the power-law behavior, we must also determine a fit window for the masses $m$, to which we fit the data.  To find the beginning of a fit range, we search for the smallest bare mass for which the expected physical inflection \emph{exists} in the quantity $\log F$ (\emph{e.g.} in Fig.~\ref{fig:f-example}, $\geq m_{0} = 0.02$).  This identifies the minimal bare mass at which physical behavior appears in the correlation function.  The renormalized mass $m$ corresponding to this bare mass is the beginning of the fit window.  The end of the fit window is determined by a change of inflection in the plots of binding energy versus renormalized mass.  This point is where the nonrelativistic power-law behavior has been overtaken by effects due to strong coupling at larger mass values.

\begin{figure}
    \centering
    \includegraphics[width=8.6cm]{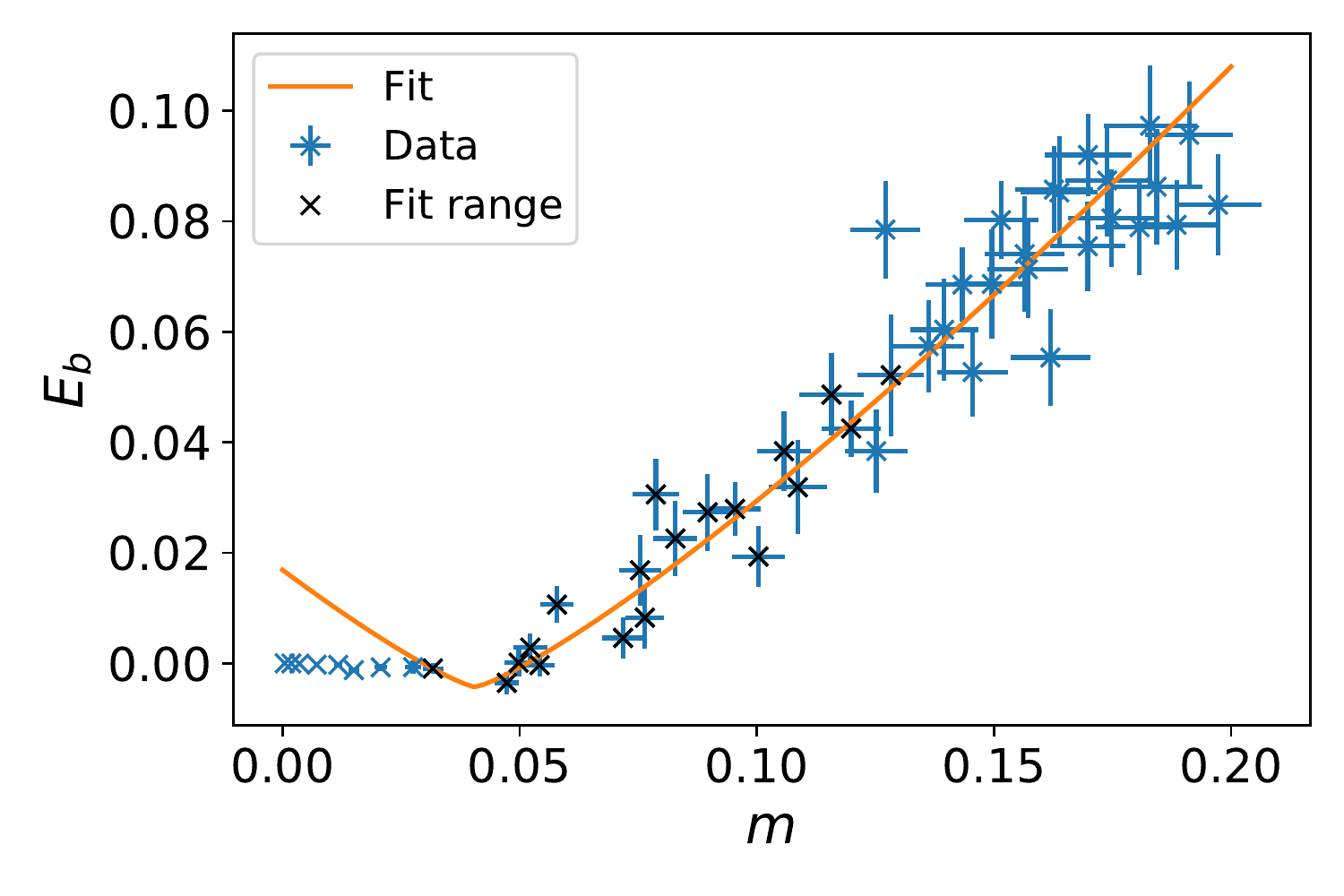}
    \caption{The power-law fit to the binding energy plotted against the renormalized mass for the $N_{4} = 4,000$, $\beta = 0.8$ ensemble.  The fit range is shown in black, and the solid line is the fit to the data.  The fit corresponds to a $\chi^{2}$/d.o.f. $= 1.24$, with a $p$-value of $0.26$.}
    \label{fig:be-rm_4kb0p8}
\end{figure}

For our power-law fits, we assume the functional form
\begin{align}
\label{eq:main-fit}
    E_{b} = A m^{\alpha}
\end{align}
where $A$ and $\alpha$ are fit parameters, which in the continuum, nonrelativistic limit are expected to be $A = G^{2}/4$, and $\alpha = 5$, as given in Eq.~\eqref{eq:newt-binding}. We find that this simple fit function is a good description of the data on all of our ensembles, with two exceptions: our two coarsest ensembles ($\beta = 1.5$ and $\beta = 0.8$).  For these ensembles we notice negative (in our convention) binding energy at small masses indicating the absence of an attractive force, which can be seen in Figs.~\ref{fig:be-rm_4kb1p5} and \ref{fig:be-rm_4kb0p8}.  We do not have a good model for how discretization effects modify the expected behavior of the binding energy at very coarse lattice spacings, but it is at least encouraging that this unphysical behavior is absent on our three finest lattice spacings.  We have the option of dropping these coarse lattices in our continuum extrapolation, and this is something we do as a cross-check, since we have to model the unphysical behavior of the binding energy on these two ensembles.  In order to describe this data, we choose a model with two additional fit parameters beyond the simple power law of Eq.~(\ref{eq:main-fit}).  The motivation for the fit function to the coarser ensembles is data driven; this is the simplest ansatz that describes the data that also reduces to the expected fit form when the new parameters are taken to zero.  Thus, for our two coarsest ensembles we use the fit function
\begin{align}
\label{eq:coarse-fit}
    E_{b} = A |x - B|^{\alpha} + C
\end{align}
with $A$, $B$, $\alpha$, and $C$ the fit parameters.  As before, $A$ and $\alpha$ can be identified with their continuum, infinite-volume counterparts in Eq.~\eqref{eq:newt-binding}.  For these two ensembles the criteria for selecting the starting mass value of the fit is never satisfied \emph{i.e.} the correct inflection in $\log[F]$ is not observed for any bare mass.  This is most likely due to large discretization errors on these coarse lattices masking the physical behavior.  Therefore the start of the fit window is somewhat arbitrary on these ensembles.  We choose a fit range that begins in the region where the binding energy trends negative and ends before the inflection in the $E_{b}$ versus $m$ plot at larger masses.  We vary this fit range to include a systematic error due to this choice.

The form in Eq.~\eqref{eq:main-fit}---even at finite lattice spacing---is re-enforced by the existence of the shift symmetry, which ensures that the bare mass is only multiplicatively renormalized, and hence, the binding energy is strictly proportional to some power of the renormalized mass.  
In Figs.~\ref{fig:be-rm_16kb-0p776}, \ref{fig:be-rm_4kb-0p6}, and~\ref{fig:be-rm_16kb0} we show examples of the binding energy plotted against the renormalized mass, along with a best fit line and the fit range used (in black), for three different lattice spacings.  These are the finer lattices at $\ell_{\rm rel} = 0.7$ and $\ell_{\rm rel} = 0.8$, and one of the coarser ensembles at $\ell_{\rm rel} = 1$, respectively.  In Figs.~\ref{fig:be-rm_4kb1p5} and~\ref{fig:be-rm_4kb0p8} we show the same quantities for the extra coarse, $\ell_{\rm rel} = 1.59$ ensemble, and $\ell_{\text{rel}} = 1.28$ ensemble.  We see good agreement between the fit functions Eq.~\eqref{eq:main-fit} and \eqref{eq:coarse-fit}, and the data.

These fits are done by taking the correlations in the data into account.  We use weighted orthogonal distance regression~\cite{odr-orig,odr-ref} to incorporate correlations in the renormalized mass and in the binding energy data sets simultaneously.  For the weights we use the inverse covariances in both data sets to obtain the best fit to the data points using $\chi^2$ minimization.  A detailed discussion of this procedure can be found in appendix~\ref{app:correlated-fit}.
To assess a systematic error associated with the choice of fit range, we vary the start and end points of a fit range over a reasonable set of values guided by the quality of fit and tabulate the results.  We then calculate the standard deviation of those results and include it as a systematic error, adding it in quadrature to the statistical error of the result from the central fit to give a total error.

We perform this power-law fit across all of our ensembles, extracting a power $\alpha$ and a coefficient $A$.  From that coefficient $A$ we calculate $\sqrt{4 A}$, which we associate with a value for $G$ at fixed volume and lattice spacing.  While this association with $G$ at finite lattice spacing or volume may suffer from systematic errors, in the continuum, infinite volume limit the quantity $\sqrt{4 A}$ should extrapolate to $G$.
With these results for $\alpha$ and $G$ across ensembles, we are able to obtain their continuum, infinite-volume values.  

\subsection{Continuum, infinite volume extrapolation}
\label{subsec:con-lim}
To preform the extrapolation to the continuum, infinite-volume limit we take the simplest ansatz suggested from finite-size scaling, and the discretization dependence suggested by the symmetries of the theory.  This approach is similar to what has been used in our previous analyses on dynamical triangulations, with the choice of fit functions given by
\begin{align}
\label{eq:alpha-fit}
    \alpha = \frac{H_{\alpha}}{V} + I_{\alpha} \ell_{\text{rel}}^{2} + \frac{J_{\alpha}}{V^{2}} + K_{\alpha} \ell_{\text{rel}}^{4} + L_{\alpha}
\end{align}
and
\begin{align}
\label{eq:G-fit}
    G = \frac{H_{G}}{V} + I_{G} \ell_{\text{rel}}^{2} + \frac{J_{G}}{V^{2}} + K_{G} \ell_{\text{rel}}^{4} + L_{G},
\end{align}
where $H_{i}$, $I_{i}$, $J_{i}$, $K_{i}$, and $L_{i}$ are fit parameters for their respective quantities.  Here $V$ is the physical volume, and $\ell_{\rm rel}$ is the relative lattice spacing.  We include quadratic corrections in the inverse physical volume, and the squared lattice spacing, since we find curvature in our data when we include small volumes and coarse lattice spacings.  In addition to the fit ansatzes in Eqs.~\eqref{eq:alpha-fit} and~\eqref{eq:G-fit}, we also perform fits dropping the $\sim \ell_{\text{rel}}^{4}$ term, which we are able to do when we also drop the two coarsest lattice spacings.  The results from this fit are consistent within one-sigma to the results of the fit to the full data set including the $\ell_{\text{rel}}^{4}$ term.

\begin{figure}
    \centering
    \includegraphics[width=8.6cm]{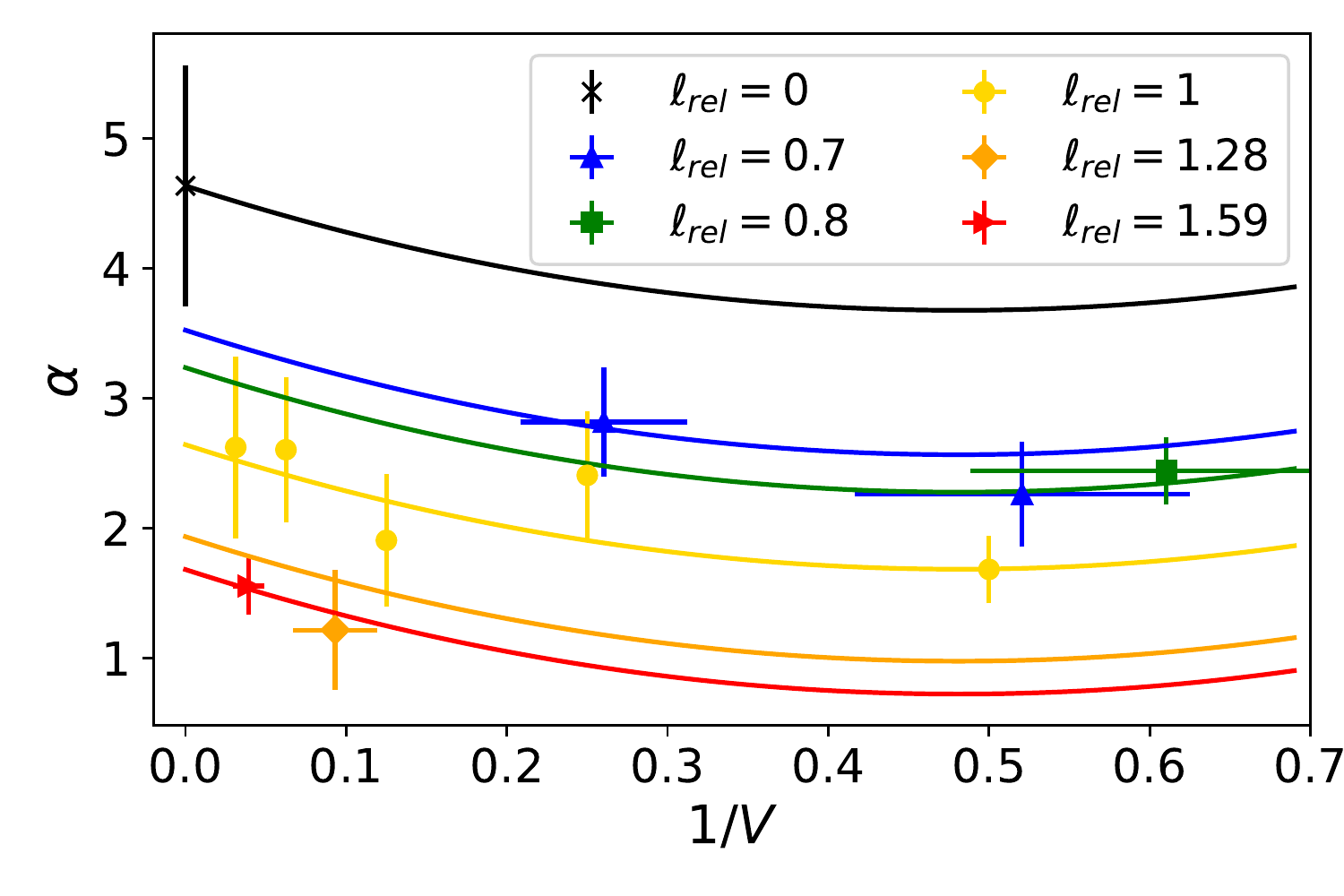}
    \caption{The power $\alpha$ as a function of the inverse physical volume (expressed in units of 1000 simplices) for all of the ensembles (colored), as well as the continuum limit (in black).  Here quadratic corrections in $1/V$ as well as $\ell_{\text{rel}}^{2}$ were used to  model the extrapolation.  For this fit we find $\chi^{2}/\text{d.o.f.} = 0.56$ corresponding to a $p$-value of 0.73, and the continuum, infinite volume value is $\alpha = 4.6(9)$.}
    \label{fig:alpha_volume}
\end{figure}

\begin{figure}
    \centering
    \includegraphics[width=8.6cm]{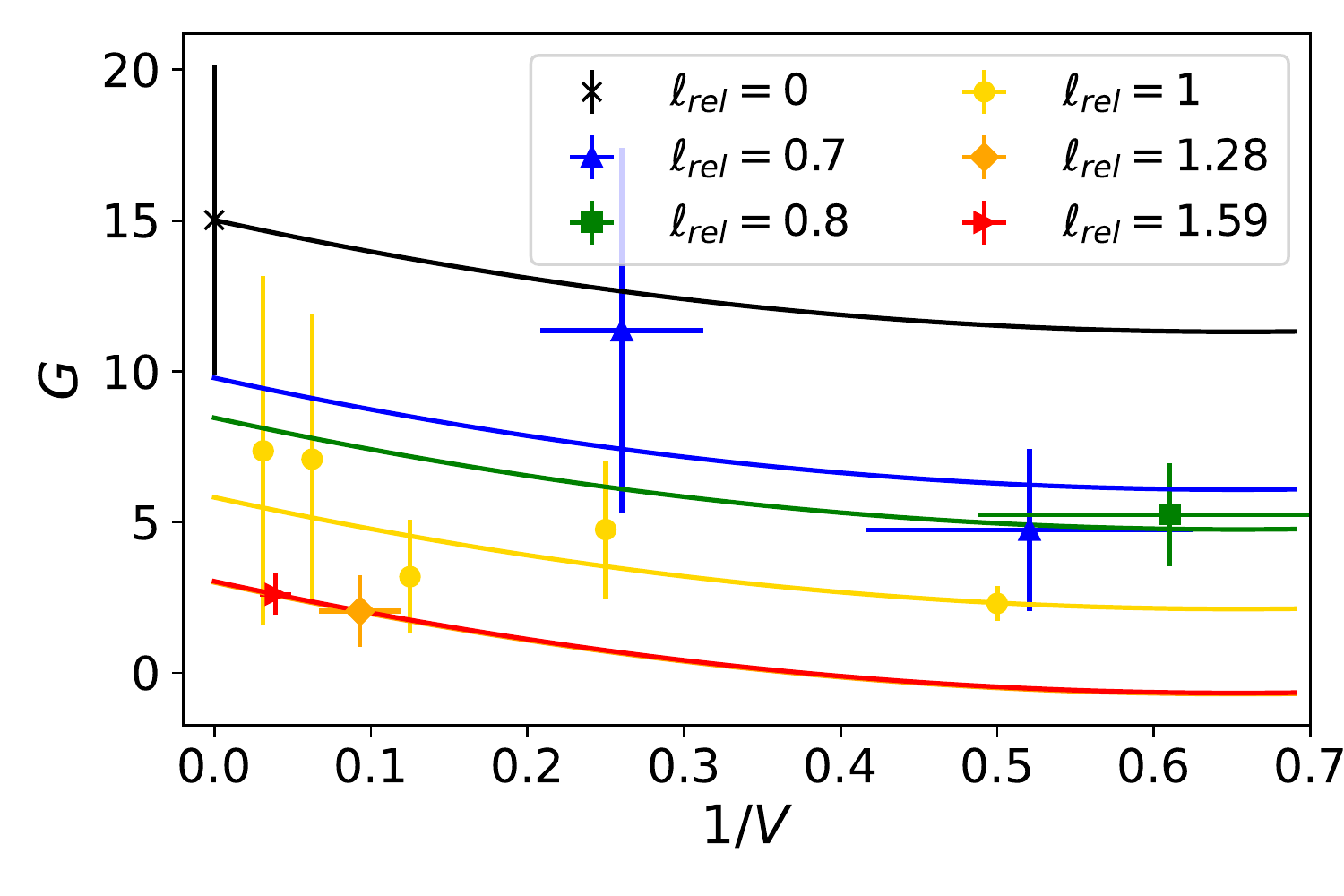}
    \caption{Newton's constant $G$ as a function of the inverse physical volume (expressed in units of 1000 simplices) for all of the ensembles (colored), as well as the continuum limit (in black).  Here quadratic corrections in $1/V$ as well as $\ell_{\text{rel}}^{2}$ were used to  model the extrapolation.  For this fit we find $\chi^{2}/\text{d.o.f.} = 0.37$ corresponding to a $p$-value of 0.87, and the continuum, infinite volume value is $G = 15(5)$.}
    \label{fig:G_volume}
\end{figure}

The extrapolations for $\alpha$ and for $G$ are shown in Figs.~\ref{fig:alpha_volume} and~\ref{fig:G_volume}, respectively, where they are plotted against the inverse physical volume.  There, lines of constant lattice spacing are drawn, and the zero-lattice-spacing limit is shown in black. 
\begin{figure}
    \centering
    \includegraphics[width=8.6cm]{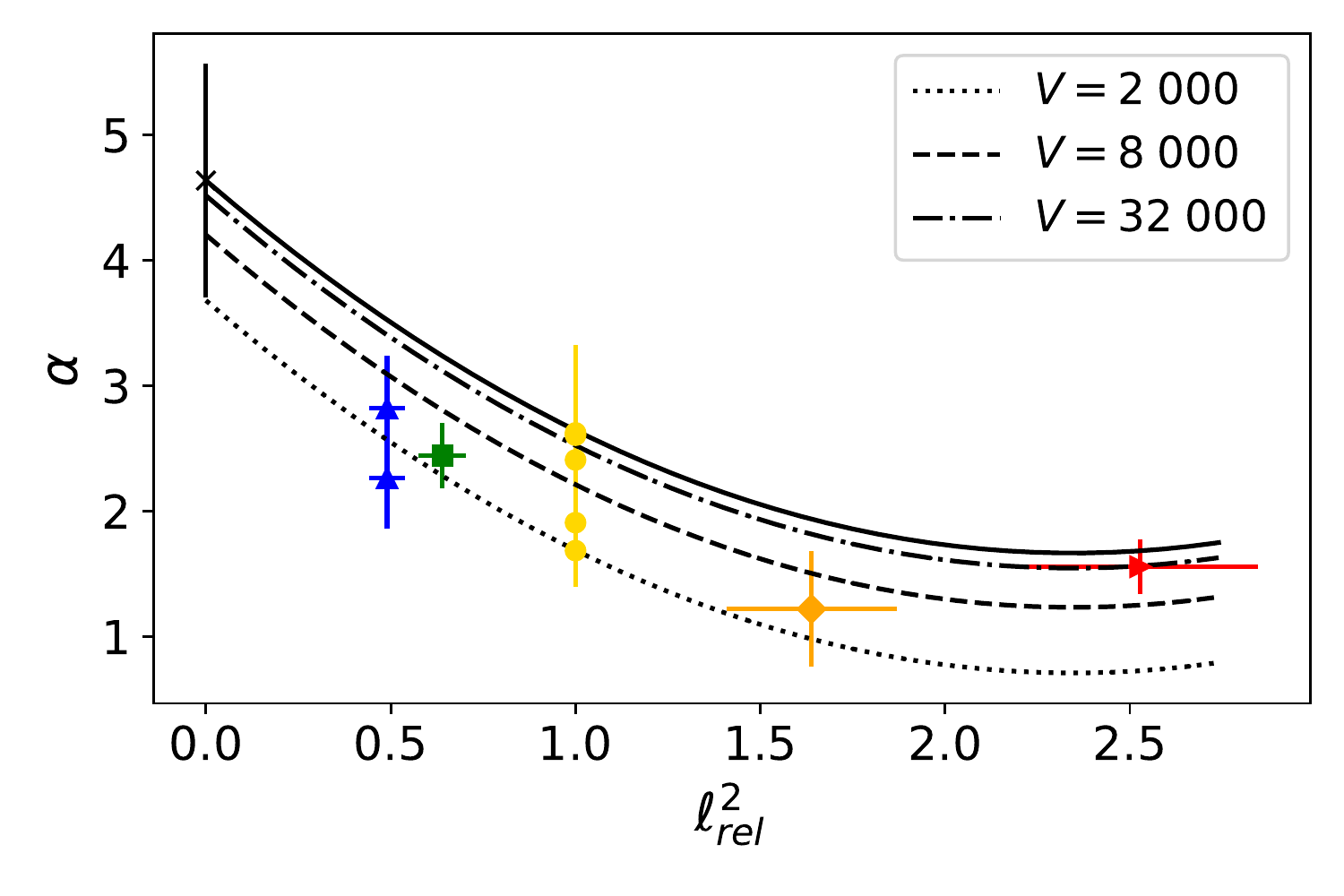}
    \caption{The same data and fit from Fig.~\ref{fig:alpha_volume} however now plotted as a function of the squared lattice spacing.  Here example lines of constant physical volume are plotted along with the infinite volume limit as a solid black line, and the data are represented in the same manner as Fig.~\ref{fig:alpha_volume}.}
    \label{fig:alpha_spacing}
\end{figure}
\begin{figure}
    \centering
    \includegraphics[width=8.6cm]{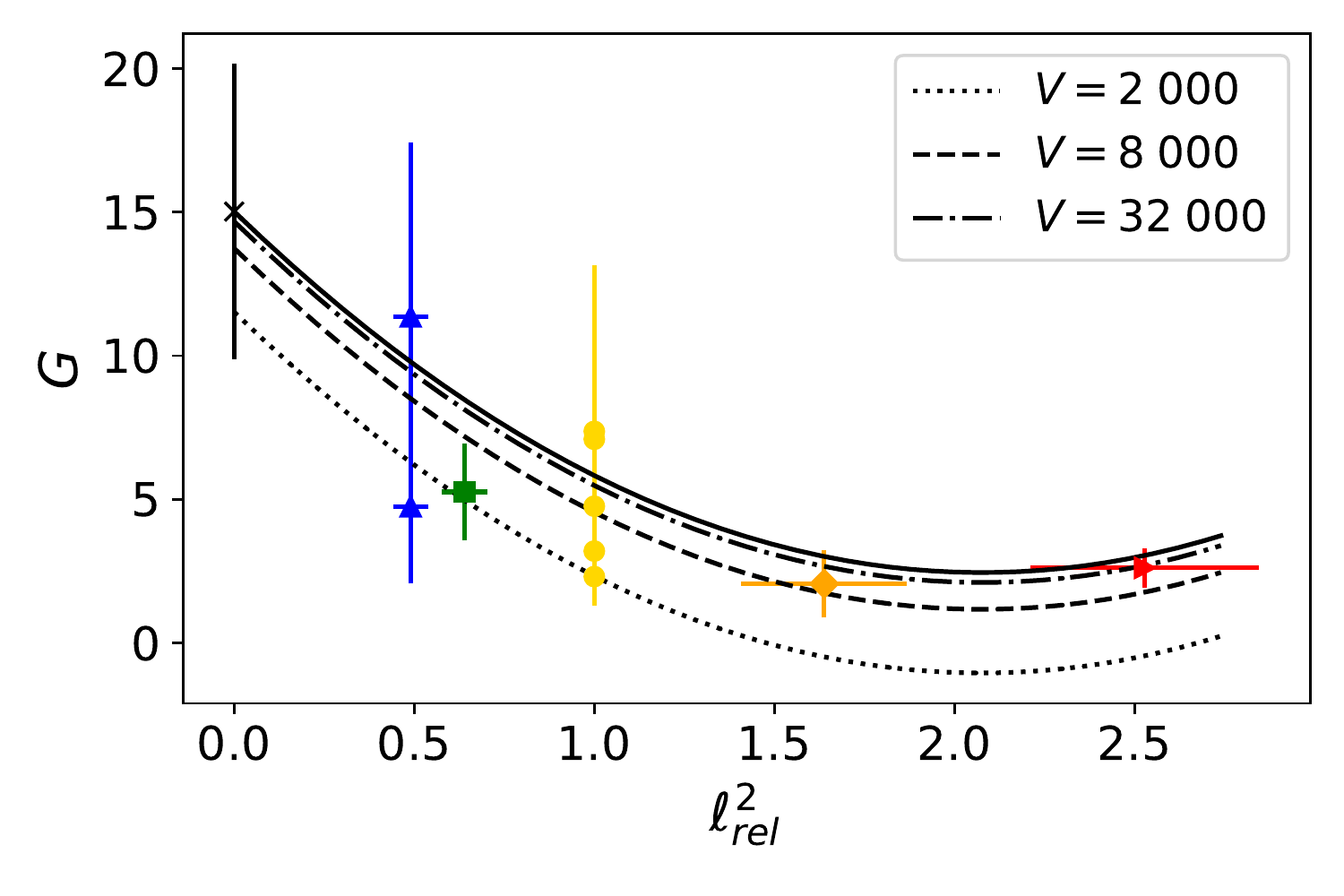}
    \caption{The same data and fit from Fig.~\ref{fig:G_volume} however now plotted as a function of the squared lattice spacing.  Here example lines of constant physical volume are plotted along with the infinite volume limit as a solid black line, and the data are represented in the same manner as Fig.~\ref{fig:G_volume}.}
    \label{fig:G_spacing}
\end{figure}
In Figs.~\ref{fig:alpha_spacing} and~\ref{fig:G_spacing} we show the same extrapolations to the infinite-volume, continuum limit as the fits shown in Figs.~\ref{fig:alpha_volume} and~\ref{fig:G_volume}, but with a different cross-section through the two-dimensional parameter space.  In these plots we show the values for $\alpha$ and $G$ plotted versus the squared lattice spacing.  There, lines of constant physical volume are drawn with the infinite-volume limit shown as a solid black line.  Note that the continuum, infinite-volume limit is taken separately for $\alpha$ and $G$.

These fits are performed assuming that the data points are uncorrelated, which is reasonable given that the points are all from different ensembles.  For the $\alpha$ extrapolation we find the $\chi^{2}/\text{d.o.f} = 0.56$, corresponding to a $p$-value of 0.73.  For the $G$ extrapolation we find $\chi^{2}/\text{d.o.f} = 0.37$, corresponding to a $p$-value of 0.87.  The infinite volume, continuum extrapolated values are $\alpha = 4.6(9)$, and $G = 15(5)$.  The errors on these quantities are fairly large, around $20\%$-$30\%$ but the value of $\alpha$ is consistent with what is needed to recover the Newtonian limit.  In fact, this result is somewhat better than it at first appears because of the sensitive dependence of $\alpha$ on the effective dimension.  Given Eq.~(\ref{eq:alpha}) for the dependence of $\alpha$ on dimension, one finds that our $1\sigma$ error band on $\alpha$ implies a $1\sigma$ error range for the dimension of between $3.6$ and $4.1$.  Thus our result implies Newtonian binding in a dimension quite close to 4 at long distance scales in the continuum limit.  Note that at coarser coupling and smaller lattice volumes where the effective dimension of the lattice geometries is around 3 or lower, the value of $\alpha$ drops to 2 or lower, which is what we expect given the dependence of $\alpha$ on dimension.  Thus our results are consistent with expectations.

The value of $G$ that we should expect is not known \emph{a priori}, since it sets the lattice spacing in physical units, but it should satisfy certain consistency checks, as discussed in Sect.~\ref{sec:newton}.  The constraint that the binding be nonrelativistic implies that $G m^2/2~\ll~1$, which translates into $m^2 \ll 0.13$ in our fiducial lattice units, given our value of $G$ in those units.  The upper range of masses in our fit window is at $m^2 \approx 0.02$, so the nonrelativistic condition is satisfied.  

One particularly nice feature of this calculation is that our value of $G$ allows us to determine the lattice spacing in units of the Planck length for the first time.  Because our extrapolation procedure recovers the correct Newtonian limit, we can have some confidence that the value of $G$ computed via this method is reliable.  We find that $\sqrt{G}=\ell_{Pl}=(3.9 \pm 0.7) \ell_{\rm fid}$.  Thus, our fiducial lattice spacing is around 1/4 the Planck length.  Our finest lattice spacing is around 1/6 the Planck length, so we see that the lattice spacing can indeed be made smaller than the Planck length.

\section{discussion \& conclusion}
\label{sec:conclusion}

One of the tests that any formulation of lattice gravity must pass is that it must have the correct classical limit.  In this work we have shown that the interaction of scalar particles is well-described by Newton's potential in the appropriate non-relativistic, classical limit.  This conclusion comes from our study of the binding energy of the two particle bound state as a function of the constituent scalar particle mass.  The analysis makes use of a number of ensembles with multiple volumes and lattice spacings, allowing us to extrapolate our results to the continuum, infinite volume limit.  

Our calculation passes a number of non-trivial cross-checks.  We show numerically that the renormalized scalar mass approaches zero as the bare mass approaches zero, which is expected given the shift symmetry of the lattice action.  We study the two-particle binding energy as a function of its constituent mass, and we find that it is well-described by a power law in the non-relativistic limit.  At finite lattice spacing and finite volume the exponent in the power law is close to what one expects if the effective dimension of the geometry is near the measured values for the effective dimension on these same lattices in Ref.~\cite{Laiho:2016nlp}.  Only in the continuum, infinite volume limit does the power law correspond to that of the Newtonian potential in four dimensions.  The consistency of the binding with Newtonian gravity allows us to extract a value of $G$ from the calculation, and  knowing $G$ allows us to determine the Planck scale for the first time within EDT.  We find that the Planck length is $\ell_{Pl}=(3.9 \pm 0.7) \ell_{\rm fid}$, so that our fiducial lattice spacing is about 1/4 the Planck length, and our finest lattice spacing is around 1/6 the Planck length.  This shows that as the continuum limit is approached the lattice spacing becomes a decreasing fraction of the Planck length, suggesting that there is no barrier to taking a continuum limit.

Although the calculation is done in the quenched approximation, such that the back-reaction of the scalar field on geometry is neglected, we still expect the binding of particles to be governed by tree-level graviton exchange.  The effects of quenching only appear at one-loop in the low-energy effective theory, so they should not affect the recovery of the classical limit, as seen here. 

Our recovery of the Newtonian potential within EDT is a strong cross-check of all of the ingredients that go into the formulation of lattice gravity used here, from our determination of the relative lattice spacing to our approach to taking the continuum, infinite-volume limit.  This non-trivial result provides powerful evidence that EDT is in fact a theory of gravity, and that investigations of the formulation as a possible realization of the asymptotic safety scenario should continue.

\begin{acknowledgments}
The authors thank Claude Bernard and Simon Catterall for valuable discussions, and we thank Simon Catterall and Fleur Versteegen for comments on the manuscript.  JL was supported by the U.S. Department of Energy (DOE), Office of Science, Office of High Energy
Physics under Award Number DE-SC0009998.  JUY was supported by the U.S. De-
partment of Energy grant DE-SC0019139, and by Fermi
Research Alliance, LLC under Contract No. DE-AC02-
07CH11359 with the U.S. Department of Energy, Office
of Science, Office of High Energy Physics. MS is supported by the German Academic Scholarship Foundation and gratefully acknowledges hospitality at Syracuse University and at CP3-Origins, University of Southern Denmark, during various stages of this project. Computations were performed in part on the Syracuse University HTC Campus Grid and were supported by NSF award ACI-1341006.  Computations for this work were also carried out in part on facilities of the USQCD Collaboration, which are funded by the Office of Science of the U.S. Department of Energy.
\end{acknowledgments}

\appendix

\section{Correlated fitting}
\label{app:correlated-fit}

When fitting the binding energy versus the renormalized mass, both quantities come with errors, and the different data points are actually correlated, both in the $x$ and $y$ data-sets.  This is because the fits for the renormalized masses and binding energies---while at different bare input masses---used the same configurations for each fit, thus introducing correlations between fits.  In order to account for this, we did simultaneous correlated fits in both the $x$ and $y$ data sets using orthogonal distance regression.  Orthogonal distance regression  attempts to minimize the radial distance between the fit function and the input data---as opposed to the vertical distance between the fit function and the $y$ data that is typical in non-linear least squares fitting.

The idea is the following: One seeks to minimize the function,
\begin{align}
    &Q(\{ \bar{x} \}, \{ \beta \}) = \\
    &\sum_{i,j} (f(\bar{x}_i, \beta_k) - y_i) w_{ij}^{-1} (f(\bar{x}_j, \beta_k) - y_j) \, +  \\
      & (\bar{x}_i - x_i) \epsilon_{ij}^{-1} (\bar{x}_j - x_j)
\end{align}
where $\bar{x}_{i}$ is the collection of ideal $x$ values which minimizes the expression, $f$ is the fitting function, $\beta_{k}$ is the collection of fit parameters, $y_i$ is the collection of $y$ data, $x_i$ is the collection of $x$ data, $w^{-1}$ is the inverse of the covariance matrix for the $y$ data, and $\epsilon^{-1}$ is the inverse of the covariance matrix for the $x$ data.  The problem can be recast, with a change of variables, into a different expression to be minimized,
\begin{align}
    & Q(\{ \delta \}, \{ \beta \}) = \sum_{i,j} \\
    &  (f(x_i + \delta_i, \beta_k) - y_i) w_{ij}^{-1} (f(x_j + \delta_j, \beta_k) - y_j) \, +  \\
      & \delta_{i} \epsilon_{ij}^{-1} \delta_j .
\end{align}
Now, there are two sets of parameters to minimize, the $\delta_i$, and the $\beta_i$, where $\delta_{i} = \bar{x}_{i} - x_{i}$ is the residual in the $x$ data.

The above expression can be recast into a very simple form for numerical purposes.  The $w^{-1}$ and $\epsilon^{-1}$ can be combined into a single block-diagonal matrix,
\begin{equation}
    C^{-1} = w^{-1} \oplus \epsilon^{-1},
\end{equation}
and the residuals can be combined into a single vector,
\begin{equation}
    v = (f(x + \delta, \beta_k) - y) \oplus \delta .
\end{equation}
This allows the above function to be written as,
\begin{equation}
    Q(\{\delta\}, \{\beta\}) = \sum_{i,j} v_{i} C^{-1}_{ij} v_{j}.
\end{equation}
For this work, the covariance matrices are positive definite, so we can take the above expression one step further and perform a Cholesky decomposition,
\begin{equation}
    C_{ij}^{-1} = \sum_{k} L_{ik} L^{T}_{kj}
\end{equation}
which in turn allows us to express $Q$ as a simple dot product between vectors, $Q = \sum_{i} r_{i} r_{i}$, with $ r_{i} = \sum_{k}  v_{k} L_{ki}$.  To minimize $Q$ we use the \emph{Levenberg-Marquardt algorithm}, and minimize with respect to the $\delta_i$ and the $\beta_i$.  The resulting $\beta$ values are the desired fit parameters for the input function, $f$.

%

\end{document}